\documentclass[conference]{llncs}

\usepackage{makeidx}  
\usepackage{algorithm}
\usepackage{graphicx}
\usepackage{wrapfig}
\usepackage{epsfig}
\usepackage{proof}
\usepackage{tikz}
\usetikzlibrary{shapes,arrows}
\usetikzlibrary{trees,positioning}
\usepackage{amsmath}
\usepackage{amssymb}
\usepackage{color}
\usepackage{lscape}
\usepackage{multirow}
\usepackage{rotating}
\usepackage{url}
\usepackage{subfigure}
\usepackage{appendix}
\usepackage{listings}
\usepackage{subfig}

\newcommand{\theCA}{\ensuremath \mathtt{CA}}
\newcommand{\Ed}{\ensuremath \mathtt{Ed}}
\newcommand{\Helen}{\ensuremath \mathtt{Helen}}
\newcommand{\CRep}{\ensuremath \mathtt{CRep}}

\newcommand{\canstoredoc}{\ensuremath \mathsf{cans}}
\newcommand{\ishead}{\ensuremath \mathsf{ish}}
\newcommand{\isemployee}{\ensuremath \mathsf{ise}}

\newcommand{\know}{\ensuremath \mathsf{knows}}
\newcommand{\msg}{\ensuremath \mathsf{msg}}

\newcommand{\principal}{\ensuremath \mathit{Principal}}
\newcommand{\infon}{\ensuremath \mathit{Infon}}
\newcommand{\attribute}{\ensuremath \mathit{Attribute}}
\newcommand{\speech}{\ensuremath \mathit{Speech}}

\newcommand{\atoi}{\ensuremath \mathsf{a2i}}
\newcommand{\stoi}{\ensuremath \mathsf{s2i}}

\newcommand{\knowzero}{\ensuremath \mathsf{uknows}}
\newcommand{\said}{\ensuremath \mathsf{said}}
\newcommand{\tdOn}{\ensuremath \mathsf{tdOn}}

\newcommand{\CRO}{CRO}

\newcommand{\acs}{\ensuremath \mathcal{ACS}}

\renewcommand{\int}{\ensuremath {\mathcal I}}

\begin{document}

\title{
Automated Analysis of Scenario-based Specifications of Distributed
Access Control Policies with Non-Mechanizable Activities \thanks{The work presented in this paper was partially supported by the FP7-ICT-2009-5 Project no.~257876,
 ``SPaCIoS: Secure Provision and Consumption in the Internet of
 Services'', and by the ``Automated Security Analysis of Identity and
 Access Management Systems (SIAM)'' project funded by Provincia
 Autonoma di Trento in the context of the ``Team 2009 - Incoming''
 COFUND action of the European Commission (FP7).}
\\
(Extended Version)} 

\author{Michele Barletta$^1$, Silvio Ranise$^2$, Luca Vigan\`o$^1$} 

\institute{$^1$Dipartimento di Informatica, Universit\`a di Verona, Italy
\email{\{michele.barletta,luca.vigano\}@univr.it} \\
$^2$FBK-Irst, Trento, Italy \\
\email{ranise@fbk.eu} }

\maketitle

\begin{abstract}
  The advance of web services technologies promises to have
  far-reaching effects on the Internet and enterprise networks
  allowing for greater accessibility of data.  The security challenges
  presented by the web services approach are formidable.  In
  particular, access control solutions should be revised to address
  new challenges, such as the need of using certificates for the
  identification of users and their attributes, human intervention in
  the creation or selection of the certificates, and (chains of)
  certificates for trust management.  With all these features, it is
  not surprising that analyzing policies to guarantee that a sensitive
  resource can be accessed only by authorized users becomes very
  difficult.  In this paper, we present an automated technique to
  analyze scenario-based specifications of access control policies in
  open and distributed systems.  We illustrate our ideas on a case
  study arising in the e-government area.
\end{abstract}

\section{Introduction}

Access control aims at protecting data and resources against
unauthorized disclosure and modifications while ensuring access to
authorized users. An access control request consists of a
\emph{subject} asking to perform a certain \emph{action} on an
\emph{object} of a system. A set of \emph{policies} allows the system
to decide whether access is granted or denied by evaluating some
conditions on the attributes of the subject, the object, and the
environment of the system (such as the identity, role, location, or
time). For centralized systems, identifying subjects, objects, and the
values of the attributes is easy since both subjects and objects can
be adequately classified by identifiers that are assigned by the
system itself. For open and distributed systems such as those based on
web technology, the situation is more complex as web servers receive
and process requests from remote parties that are difficult to
identify and to bind with their attribute values. Hence,
\emph{certificates} or credentials, attesting not only the identity
but also the attributes of parties, must be exchanged to correctly
evaluate access control queries. In many situations, the creation and
exchange of certificates require human intervention, e.g., to issue
and sign a certificate or to pick one in a portfolio of available
credentials. Furthermore---as observed
in~\cite{beyond-proof-compliance} among others---in distributed
systems, a certificate can be accepted or rejected depending on the
\emph{trust relation} between the receiver and its issuer. Additional
flexibility can be gained by chains of credentials and trust. In this
context, guaranteeing that only trusted users can access sensitive
resources becomes a daunting task.

\subsection{Main Contributions} 

In this paper, we propose a technique for the automated analysis of
access control systems (ACS) in presence of human activities for the creation
and exchange of certificates together with trust management. Our
approach combines a logic-based language with model checking based on
\emph{Satisfiability Modulo Theories (SMT)} solving. More precisely, we
follow~\cite{constraintdatalog} and use \emph{Constraint Logic
Programming (CLP)} for the specification of policies and
trust management with ideas adapted from~\cite{GurevichNeeman-dkal}. The
exchange of certificates and their interplay with the set of policies
is modeled as a transition system of the type proposed
in~\cite{lietal2005}. We show that interesting
analysis problems of ACSs can be reduced to reachability problems.
Our main contribution is a decidability result for the (bounded)
reachability problem of a sub-class of transition systems that can
encode the analysis of scenario-based specifications of ACSs,
 i.e.\ situations in which the exchange of certificates is
constrained by a given causality relation. Another contribution is a
technique to reduce the number of possible interleavings while visiting
reachable states.

\subsection{A Motivating Example: the Car Registration Office} 
\label{sec:runningexe}

We consider a simplified version of the
\emph{Car Registration Office} (\CRO{}) application in~\cite{BRV-TR09}.
It consists of a citizen wishing
to register his new car via an on-line service provided by the \CRO{}.  
An employee of the \CRO{}, $\Ed$, checks if the request can be accepted according to
some criteria. If so, $\Ed$ must store the request in a central
repository $\CRep$, which, in turn,
checks if $\Ed$ is entitled to do so. 
To be successful, the storage request must be supported by three certificates: $\isemployee$ saying
that $\Ed$ is an employee of the \CRO{}, $\ishead$ saying that $\Helen$ is the head of the \CRO{} and $\canstoredoc$ saying that $\Helen$ granted $\Ed$ the permission to store documents
in $\CRep$. 
\begin{figure}[t]
\centering
  \includegraphics[scale=.5]{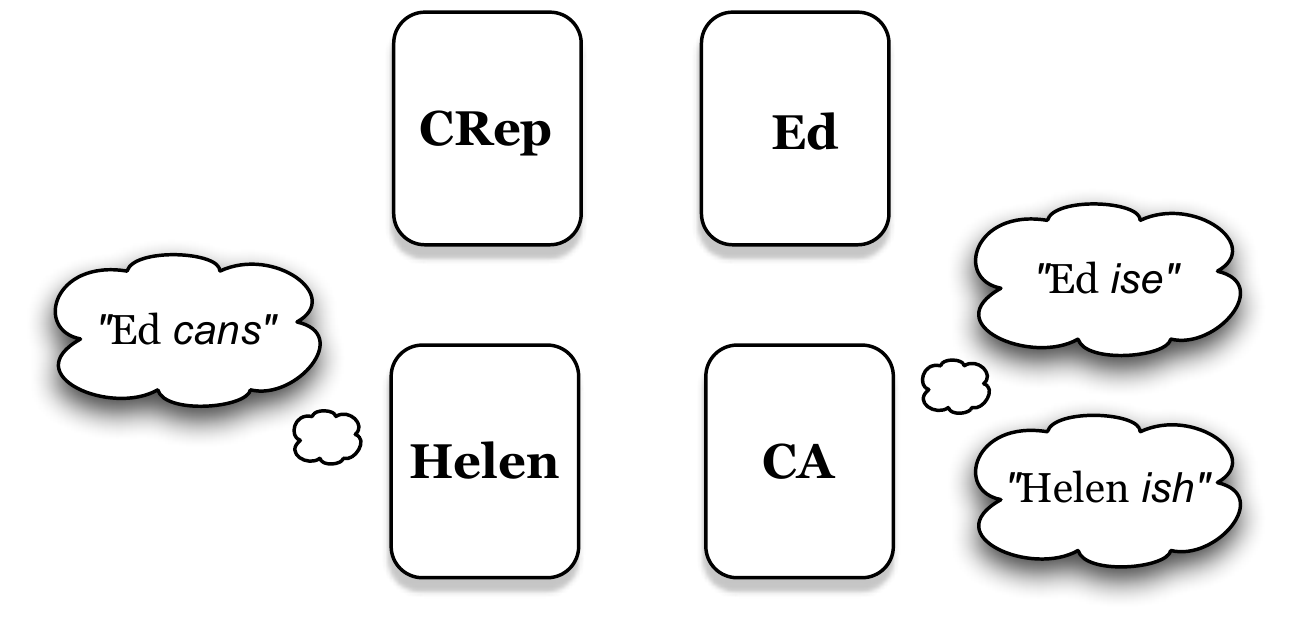}
  \caption{\label{fig:1}The \CRO{} scenario} 
\end{figure}
Roles certificates must be signed by a
trusted Certification Authority ($\theCA$) while $\Ed$'s
permission certificate is signed by $\Helen$; if these were not the case, the
certificates should be rejected because the principal that signed the
properties is \emph{untrusted}. The generation of certificates
(depicted in Fig.~\ref{fig:1}) is a \emph{non-mechanizable} activity
whose execution depends on decisions that are not modeled in the
system but only on the human behavior.  Another issue is how the
certificates are sent to $\CRep$ in order to support $\Ed$'s storage
request.  It can be $\Ed$ to send the certificates along with the
request (user-pull) or it can be $\CRep$ to collect the necessary
certificates upon reception of $\Ed$'s request (server-pull).

\subsection*{Organization of the paper} 
In Section~\ref{sec:ex-overview}, we give an
overview of the main features of our approach, which we then detail in
Section~\ref{sec:DKAL-light} where we
formalize a class of access control schemas.  In Sections~\ref{sec:symb-exec} and \ref{sec:verification}, we present our main contributions: an automated analysis technique of scenario-based specifications and a heuristics for its scalability.
In Section~\ref{sec:conclusions}, we conclude and discuss related  work.
Formal preliminaries, a complete derivation of the \CRO{} main query and an implementation of the scenario by using DKAL language, can be found in the appendix Section \ref{appx}.

\section{Overview of the Main Features of our Approach}
\label{sec:ex-overview}
Our goal is to automatically analyze situations in which (a)
certificates are created or exchanged because of human intervention, (b)
there is a need to reason about chains of credentials to establish the
validity of a certificate, and (c) message exchanges comply with a
causality relation.
\subsection{Certificates and non-mechanizable activities} 
Inspired by~\cite{constraintdatalog,GurevichNeeman-dkal}, we use a variant of
Constraint Logic Programming (CLP) to abstractly represent certificates
as well as to specify and reason about the trust relationships among
principals and the restrictions on delegating the ability to issue
certificates. 
\begin{example}
\label{ex:certsf}
For the \CRO{} scenario, the three certificates depicted
in Fig.~\ref{fig:1} can be expressed as the following CLP facts: 
\begin{align*}
(F1)\quad & \knowzero(\theCA, \atoi(\Ed,\isemployee)) \\
(F2)\quad & \knowzero(\theCA, \atoi(\Helen,\ishead))\\
(F3)\quad & \knowzero(\Helen, \atoi(\Ed,\canstoredoc)), 
\end{align*}
where $\knowzero$ represents the knowledge of
a principal resulting from non-mechanizable activities only,
called \emph{internal} knowledge, and $\atoi$ is a constructor for the
piece of knowledge about the binding of a property (e.g., being an
employee, $\isemployee$) with a principal (e.g., $\Ed$).
\qed
\end{example}

\subsection{Exchange of certificates among principals} 
\label{subsec:exchangecerts}
Distributed access control is based on exchanging  certificates among principals
so that access decisions can be taken by one principal with all the necessary information. 
So, we need to specify the actions that change the state of the system,
that is the content of the network and the internal knowledge of the principals involved. 
To this end, we use the notion of transition system introduced in~\cite{lietal2005} for
access control systems as follows. The network of messages is modeled by
a ternary predicate $\msg$ with three arguments: the sender, the payload, and
the receiver of the message. The action of $p$ sending a message with
payload $x$ to $q$ can be written as a transition
\begin{eqnarray}
  \label{eq:send-action}
  \know(p,x) \Rightarrow \oplus\msg(p,\said(x),q)
\end{eqnarray}
where $\know$ represents the knowledge of a principal, both
internal and acquired from the reception of messages from other
principals, and $\said$ transforms a piece of knowledge into an
assertion that can be communicated to other principals.  The fact that
internal knowledge is knowledge can be expressed by the 
CLP rule
\begin{eqnarray}
  \label{eq:internal-knowledge-is-knowledge}
  \know(p,x) \leftarrow \knowzero(p,x)
\end{eqnarray}
and the action of $q$ receiving a message from $p$ with $s$ as payload
is written as
\begin{eqnarray}
  \label{eq:receive-action}
  \know(q,\stoi(p,s)) \leftarrow \msg(p,s,q)
\end{eqnarray}
where $\stoi$ is a constructor for the piece of knowledge about the
binding of the utterance $s$ with a principal $p$.  
\begin{example}
\label{ex:certs}
For example, the action of $\theCA$ sending the certificate that $\Ed$
is an employee to $\Ed$ himself can be formalized as an instance of
(\ref{eq:send-action}), the reception of such a certificate by $\Ed$
as an instance of (\ref{eq:receive-action}), and the derivation that
$\Ed$ knows that $\theCA$ has uttered (and signed) the property about
$\Ed$ being an employee---formally,
$\know(\Ed,\stoi(\theCA,\said(\atoi(\Ed,\isemployee))))$---as an
application of fact (F1) and rule
(\ref{eq:internal-knowledge-is-knowledge}).
Notice that $\Ed$ cannot claim to know that he is an employee since he
does not know whether $\theCA$ is trusted on emitting this type of
utterances.  For this, suitable trust relationships should be
specified. \qed
\end{example}

\subsection{Trust relationships among principals} 
\label{subsec:trustrelations}
We use again CLP rules.
One rule is generic while the others are application dependent. The generic rule is
\begin{eqnarray}
  \label{eq:trust-app}
  \know(p, x) & \leftarrow &
   \know(p, \stoi(q,\said(x))) \wedge 
   \know(p, \atoi(q,\tdOn(x))) 
\end{eqnarray}
saying that a principal $p$ may expand its knowledge to include the
piece of information $x$ as soon as another principal $q$ has uttered
$\said(x)$ and 
$q$ is trusted on the same piece of
knowledge $x$ (the last part is encoded by the term
$\atoi(q,\tdOn(x))$).

\begin{example}
\label{ex:rules}
In the case of the \CRO{}, we need also to consider the following four
specific CLP rules, that encode the trust relationships among the
various principals:
\begin{eqnarray*}
(P1) && \know(\CRep,\atoi(p, \canstoredoc))  \leftarrow 
       \know(\CRep, \atoi(q, \ishead)) \wedge \\
     &&
       \know(\CRep, \atoi(p, \isemployee)) \wedge 
       \know(\CRep, \stoi(q,\said(\atoi(p, \canstoredoc)))) \\
(P2) && \know(p, \atoi(\theCA, \tdOn(x))) \\
(P3) && \know(p,\atoi(q,\tdOn(\stoi(\theCA,\said(x))))) \\ 
(P4) && \know(p, \atoi(q,\tdOn(\stoi(r, \said(\atoi(q, \canstoredoc))))) ) 
       \leftarrow  \know(p, \atoi(r, \ishead)),
\end{eqnarray*}

$(P1)$ says that a principal $p$ can store documents in the $\CRep$ if he
is an employee of the \CRO{} and his head permits it, $(P2)$
says that the content of any utterance of the $\theCA$ is trusted, 
$(P3)$ says that an utterance of
a principal repeating an utterance of the $\theCA$ is trusted, and finally 
$(P4)$ says that the head of the \CRO{} is trusted when emitting an utterance
granting permission to store documents in the $\CRep$ to a principal.
\qed
\end{example}


\subsection{Automated analysis of scenarios}
The formal framework sketched above allows us to develop automated analysis techniques to verify the \emph{availability} (policies suitable for scenario's execution) or the \emph{security}  (critical operations performed by trusted principals) of typical
scenarios in which an ACS should operate.  
Availability implies that
the policies are not too restrictive to prevent the scenario to be
executable while security means that only trusted principals are
granted access to sensitive resources or perform sensitive operations.
Both problems can be reduced to check whether, after performing a
sequence of non-mechanizable activities and exchanging messages among
principals, it is possible to reach a configuration of the network in
which an access control query (e.g., in the \CRO{}, ``Can $\Ed$ store
the citizen's request in $\CRep$?'') gets a positive or a negative
answer.

In other words, we want to solve problems as stated by the following definition: 
\begin{definition}[Reachability problems] 
\label{def:reachprobs}
Given the following conditions:
\begin{itemize}
\item let the network be initially empty (formally, $\msg$ is
interpreted as an empty relation),
\item  $H_0$ be a set of facts derived
from non-mechanizable activities (e.g., (F1), (F2) and (F3) described in Example \ref{ex:certsf}),	
\item and $G$ be a conjunction of $\know$-facts
describing an access control query (e.g., $\know(\CRep,
\atoi(\Ed,\canstoredoc))$ for the \CRO{} example)
\end{itemize}
we aim to check if
does there exist a sequence of $n$ instances of the transition rule
(\ref{eq:send-action}) and a sequence $H_1, ..., H_n$ of $\know$-facts
derived from non-mechanizable activities, such that $G$ is satisfied in
the final state?
\end{definition}
To practically answer this question, initially we need to compute the fix-point of the facts in $H_0$ with the CLP rules (\ref{eq:internal-knowledge-is-knowledge}),
(\ref{eq:receive-action}), (\ref{eq:trust-app}) and those formalizing
specific trust relations.  This process must be repeated at each step
$i=1, ..., n$ with the facts describing the content of the network
(derived by applying (\ref{eq:receive-action}) at step $i-1$), those
in the set $H_i$, and the CLP rules.  Since more than one transition
(\ref{eq:send-action}) can be enabled at any given step $i$, it is
necessary, in general, to consider several possible execution paths.

Not surprisingly, the reachability problem turns out to be quite
difficult.
Fortunately, in scenarios with constrained message exchanging (e.g., the user-pull or the
server-pull configurations considered for the \CRO{} above), the
reachability problem becomes simpler. It is possible to fix a bound $n$
of transitions to consider and apply a reduction technique to decrease 
the number of different execution paths to be explored as we will see in 
Sections~\ref{sec:symb-exec} and~\ref{sec:verification}.

\section{A Class of Access Control Schemas}
\label{sec:DKAL-light}

According to~\cite{lietal2005}, we report, in the following, the definition of \emph{access control schema} (in short $\acs$).

\begin{definition}[Access Contro Schema]
\label{acss}
An $\acs$ is a transition system 
$$(S, Q, \vdash,\Psi), $$ 
where $S$ is a set of states, $Q$ is a set of queries, $\Psi$ is a set of state-change rules, and $\vdash\, \subseteq S\times Q$ is the relation establishing if a query
$q\in Q$ is satisfied in a given state $\gamma\in S$, written as
$\gamma \vdash q$. 
\end{definition}

For $s, s'\in S$ and $\psi\in \Psi$, we write $s
\rightarrow_{\psi} s'$ when the change from $s$ to $s'$ is allowed by
$\psi$. The reflexive and transitive closure of $\rightarrow_{\psi}$
is denoted by $\rightarrow_{\psi}^*$. 

Given an $\acs$ $(S, Q, \vdash, \Psi)$, an instance $(s, q, \psi)$ of the
\emph{reachability problem} (see Definition \ref{def:reachprobs}) (where $s\in S$, $q\in Q$, and $\psi\in \Psi$) consists of asking whether there exists an $s'\in S$ such that
$s \rightarrow_{\psi}^* s'$ and $s' \vdash q$.

\subsection{The substrate theory $T_{\mathsf{S}}$} 
\label{subsec:substh}
We define a class of
$\acs$s by using formulae of (many-sorted) first-order
logic~\cite{enderton} to represent states and transitions. To do this
formally, we need to introduce a \emph{substrate} theory
$T_{\mathsf{S}}$, i.e., a set of formulae that abstractly specifies the
basic data-structures and operations relevant for both access control
and trust management. The theory contains a (countably) infinite set of
constants of sort $\principal$ to identify users, suitable operations to
build $\attribute$ values, 
and the functions $\atoi:\principal \times
\attribute \rightarrow \infon$, $\stoi : \principal \times
\speech \rightarrow \infon$, $\said: \infon \rightarrow \speech$,  $\tdOn :
\infon \rightarrow \attribute$, that have been already informally described in Section \ref{sec:ex-overview}.

Moreover the substrate theory $T_{\mathsf{S}}$ contains the
predicate symbol $\mathsf{prim}:\attribute$ 
that intuitively characterizes the set of ``primitive'' attributes, i.e., those already in the substrate that are not created by the ``function'' $\tdOn$ (e.g., $\isemployee, \ishead, \canstoredoc$ for the \CRO{} example).
So, it is necessary to add to the substrate theory the  following axiom $$ \forall x, a.\,
\tdOn(x)=a\Rightarrow \neg \mathsf{prim}(a)$$ where $x$ is a variable of
sort $\infon$ and $a$ is a variable of sort
$\attribute$.\footnote{Indeed, the models of the theory considered here
are a super-class of those considered in~\cite{GurevichNeeman-dkal}.
Here, we trade precision for the possibility of designing an automated
procedure for discharging a certain class of proof obligations that
encode interesting security analysis problems for the class of access
control schemas that we are defining.} 
Another important aspect we want to remark is that, even if in this paper we assume, for the sake of simplicity,
the standard situation (see, e.g., \cite{MauSch96}) where insecure communication channels
between each pair of principals are always available, it is easy to extend the substrate theory
by adding axioms to characterize the ``topology'' of the system. 

We recall that the theory
$T_{\mathsf{S}}$ identifies a class of structures that are models of
all formulae in $T_{\mathsf{S}}$ and say that a formula $\varphi$ is
satisfiable modulo $T_{\mathsf{S}}$ iff there exists a model of
$T_{\mathsf{S}}$ that makes $\varphi$ true.

\subsection{The set $S$ of states} 
\label{subsec:setofstates}
We consider the two predicate symbols
$\knowzero : \principal \times \infon$ and $\msg : \principal \times
\speech \times \principal$ already introduced in Section \ref{sec:ex-overview}. 
We assume the availability of a finite set $\mathit{Po}$ of CLP rules, also called
\emph{policies}, of the form

\begin{eqnarray}
  \label{eq:policy-rule}
  A_0(\underline{x}) \leftarrow 
  A_1(\underline{x},\underline{y}) \wedge \cdots \wedge 
  A_n(\underline{x},\underline{y}) \wedge
  \xi(\underline{x},\underline{y}) ,
\end{eqnarray}
where $\underline{x}$ and $\underline{y}$ are tuples of variables,
$A_0$ is $\know$, $A_i\in \{ \know,\knowzero\}$ for $i=1, ..., n$, and
$\xi$ is a quantifier-free formula of the substrate theory
$T_{\mathsf{S}}$.  We assume $\mathit{Po}$ to always contain
(\ref{eq:internal-knowledge-is-knowledge}), (\ref{eq:receive-action}),
and (\ref{eq:trust-app}).  Given a set $F$ of constrained ground facts
and the set $\mathit{Po}$ of policies, the set $S$ of states contains
all the constrained ground facts obtained by computing the
least-fixpoint $\mathit{lfp}(F\cup\mathit{Po})$ of the ground
immediate consequence operator on $F\cup \mathit{Po}$ (see,
e.g., \cite{constraintdatalog}).

\subsection{The set $Q$ of queries and the satisfaction relation $\vdash$}
\label{subsec:setofqueries}
A \emph{query} is a conjunction of ground facts of the form
$\knowzero(p,x)\leftarrow \xi(p,x)$. We define $\vdash$ to be the
standard consequence relation $\models$ of first-order
logic~\cite{enderton}.

\subsection{The set $\Psi$ of state-change rules} 
\label{subset:setofstatechangerules}
A \emph{state-change
rule} is a formula of the form
\begin{multline}
  \label{eq:change-rule-formula}
  \exists p,x,q. \
    \know(p,x)\ \wedge 
    \forall y,z,w.\ \msg'(y,z,w) \Leftrightarrow \\
      \msg(y,z,w) \ \vee \
      (y=p\wedge z=\said(x)\wedge w=q)
\end{multline}
that is usually abbreviated as (\ref{eq:send-action}).  Intuitively,
the unprimed and primed versions of $\msg$ denote the state of the
network immediately before and after, respectively, of the execution
of the state-change rule.  Let $S_1$ and $S_2$ be two states in $S$
and $\psi$ be a formula of the form (\ref{eq:change-rule-formula}),
then $S_1\rightarrow_{\psi} S_2$ iff
\begin{multline*}
  S_2 \ := \
  \{ 
  \msg(y,z,w) \leftarrow 
  (y=p \ \wedge \
    z=\said(x) \ \wedge \
    w=q)\sigma 
  \mid 
    S_1\cup \{ \know(p,x)\sigma \} \text{ is } \\
    \text{ satisfiable modulo $T_{\mathsf{S}}$ for } \sigma   
    \text{ ground substitution of } p,x,q 
\}\,.
\end{multline*}
When $S_2\neq \emptyset$, the state-change rule is \emph{enabled} in
$S_1$; otherwise (i.e., $S_2= \emptyset$) it is \emph{disabled in
  $S_1$}.  This concludes the definition of our class of access
control schema. 

\subsection{Reachability problems} 
In the class of $\acs$s defined above, 
policies rely on conditions that are
determined by the exchange of messages (cf.\ predicate $\msg$ and the
CLP rule (\ref{eq:receive-action})) and non-mechanizable activities
(cf.\ predicate $\knowzero$ and the CLP rule
(\ref{eq:internal-knowledge-is-knowledge})). The state-change rules in
$\Psi$ can only modify $\msg$ and leave $\knowzero$ unconstrained since
it is very difficult to model how humans decide to
create a certain certificate. 
Returning to the \CRO{} scenario, consider the assertion of fact (F3) as an example of
a certificate that can be created at any time of the execution sequence of the system. 
To emphasize this aspect, we explicitly define the notion of (instance of)  the reachability problem, although technically it can be derived from that of  reachability problem given at the beginning of this
section.

\begin{definition}[Instance of the reachability problem]
\label{def:instofreachprob}
Given a set $\mathit{Po}$ of policies and a query $G$, an
  \textbf{instance of the reachability problem} amounts to
  establishing whether there exist an integer $n\geq 0$ and constraint (ground)
   facts $H_0(\knowzero_0), ...,$
  $H_{n-1}(\knowzero_{n-1})$ such that
  \begin{eqnarray}
    \label{eq:bmc}
    \mathcal{R}_n\cup \{ G(\know_n) \} \text{ is } \text{satisfiable modulo } 
    T_{\mathsf{S}}\,,
  \end{eqnarray}
  where $\mathcal{R}_0 := \mathit{lfp}(\{H_0(\knowzero_0)\} \cup
  \mathit{Po}(\know_0))$, $\mathcal{R}_i \rightarrow_{\psi}
  \mathit{R}_{i+1}$, $\mathcal{R}_{i+1} :=
  \mathit{lfp}(\mathit{R}_{i+1} \cup \{H_{i+1}(\knowzero_{i+1})\} \cup
  \mathit{Po}(\know_{i+1}))$, $\msg_i$, $\knowzero_{i}$, and
  $\know_{i}$ denote uniquely renamed copies of $\msg, \knowzero$, and
  $\know$, respectively, and $\alpha(s_i)$ is the formula obtained
  from $\alpha$ by replacing each occurrence of the symbol $s$ with
  the renamed copy $s_i$ (for $i=0, ..., n+1$).  
  \end{definition}

Intuitively, $\mathcal{R}_0$ is the initial knowledge of the
principals computed from their internal knowledge and the (exhaustive)
application of the policies without any exchange of messages (recall
that, initially, we assume that the network contains no messages).
Then, $\mathcal{R}_1$ is obtained from $\mathcal{R}_0$ by first
applying one of the available state-change rule ($\mathit{R}_{i+1}$)
followed by the exhaustive application of the policies that allows
each principal to possibly derive new knowledge from both the
exchanged messages and their internal knowledge.  The
$\mathcal{R}_i$'s for $i\geq 2$ can be similarly characterized.

When there exists a value of $n$ such that
(\ref{eq:bmc}) holds, we say that $G$ is \emph{reachable}; otherwise
(i.e., when, for every $n\geq 0$, we have that (\ref{eq:bmc}) does not
hold) we say that $G$ is \emph{unreachable}.  If a bound $\overline{n}$ on $n$ is
known, we talk of a \emph{bounded} reachability  problem (with
bound $\overline{n}$). 
Since the reachability problem is undecidable even without
considering non-mechanizable facts (see~\cite{BRV-TR12} for
details) in the rest of the paper, we prefer to focus on identifying restricted
instances of the (bounded) reachability problem that are useful in
practice and can be automatically solved.

\section{Automated Analysis of Scenario-based Specifications} 
\label{sec:symb-exec}
Web service technology supports the development of distributed systems
using services built by independent parties.  Consequently, service
composition and coordination become an important part of the web
service architecture.  Usually, individual specifications of web
services are complemented by scenario-based specifications so that not
only the intentions of individual services but also their expected
interaction sequences can be documented.  Interestingly, as we will
show below, scenario-based specifications can be exploited to
automatically and efficiently analyze security properties despite the
well-known fact that unforeseen interplays among individually secure
services may open security holes in their composition.  The idea is to
associate a scenario with an instance of a bounded reachability
problem and then consider only the sequences of state-change rules
that are compatible with the scenario itself.

\subsection{Scenarios and bounded reachability problem}  

\begin{figure}[t]
  \centering
  \includegraphics[scale=.6]{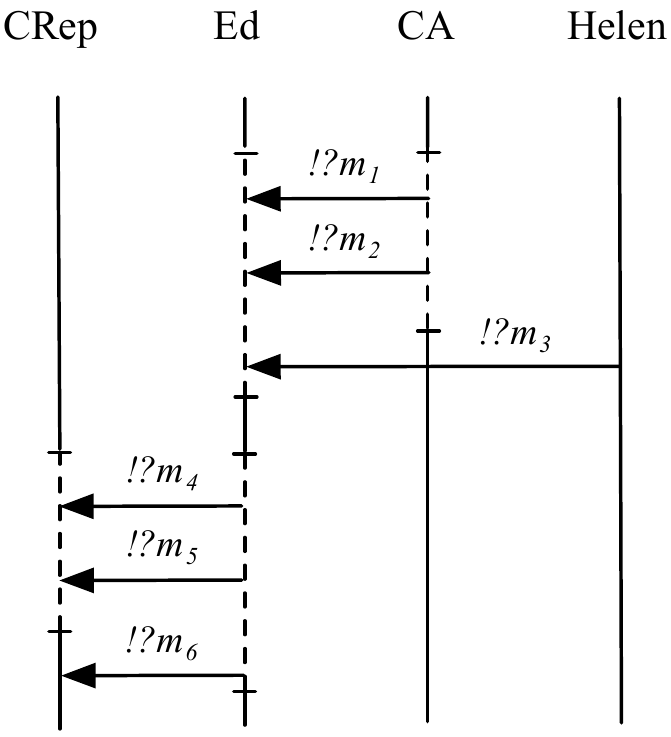} 
  \quad\quad\quad
  \raisebox{1.5cm}{%
  {
    \begin{tabular}{|c||l|}
    \hline
    $m_1$ & $\said(\atoi(\Ed,\isemployee))$ \\ \hline
    $m_2$ & $\said(\atoi(\Helen,\ishead))$  \\ \hline
    $m_3$ & $\said(\atoi(\Ed,\canstoredoc))$ \\ \hline
    $m_4$ & $\said(\stoi(\theCA, \said(\atoi(\Ed,\isemployee))))$ \\ \hline
    $m_5$ & $\said(\stoi(\theCA, \said(\atoi(\Helen, \ishead))))$ \\ \hline
    $m_6$ & $\said(\stoi(\Helen, \said(\atoi(\Ed, \canstoredoc))))$ \\ \hline
    \end{tabular}}}
  \caption{\label{fig:msc-cro}A user-pull scenario for the \CRO{}:
    $\Ed$ sends to $\CRep$ the certificates for a positive answer to
    the query $G:=\know(\CRep, \atoi(\Ed, \canstoredoc))$ }
\end{figure}

In our framework, a scenario is composed of a finite set of principals,
some sequences of state-change rules of finite length, and a query $G$
that encodes an availability or a security property. Since a
state-change rule (\ref{eq:send-action}) to be enabled requires a
principal to have some internal knowledge, this component of the
scenario implicitly identifies a sequence $H_0(\knowzero_0), ...,
H_{n-1}(\knowzero_{n-1})$ of non-mechanizable facts where $n\geq 0$ is
the length of the longest sequence of state-change rules. 

\begin{example}
An example of informal specification of a scenario is the Message Sequence Chart (MSC)
for the \CRO{} depicted on the left of Fig.~\ref{fig:msc-cro} where the
$m_i$'s are the messages containing the utterances in the table on the
right of the same figure. It is easy to find the instance of
(\ref{eq:send-action}) that allows one to send each message $m_i$. The
solid lines in the MSC impose an ordering on messages while dashed lines
(called co-regions, see, e.g., \cite{rudolph96}) do not. So, for
example, $\theCA$ can send the two certificates (in messages $m_1$ and
$m_2$) about the roles of $\Ed$ and $\Helen$ to $\Ed$ in any order and
these two certificates as well as the one sent by $\Helen$ about
granting the permission to store documents in $\CRep$ (in message $m_3$)
can be received in any order by $\Ed$. For the \CRO{}, the query
$G:=\know(\CRep, \atoi(\Ed, \canstoredoc))$ encodes an availability
property saying that the trusted user $\Ed$ can get the permission of
storing the document in $\CRep$. Since the length of sequence of
state-change rules specified by the scenario is $6$, we can build an
instance of the bounded reachability problem with bound $n=6$ and the
following sequence of non-mechanizable facts: $H_0:=\knowzero(\theCA,
\atoi(\Ed,\isemployee))$, $H_1:=\knowzero(\theCA,
\atoi(\Helen,\ishead))$, $H_2:=\knowzero(\Helen,
\atoi(\Ed,\canstoredoc))$, and $H_i:=\mathit{true}$ for $i=3,4,5$.
Other sequences are compatible with the scenario above, we just picked
one. Such sequences are finitely many and can be exhaustively enumerated.
\qed
\end{example}

\subsection{Decidability of a class of instances of the reachability problem}
 It would be interesting to find conditions that guarantee the
decidability of this kind of instances of the bounded reachability
problem with a given sequence of non-mechanizable facts. Before doing
this, we need to discuss the following four technical conditions on the
substrate theory $T_{\mathsf{S}}$.

First, 
the fact that there is a finite and known number of principals
in any scenario can be formalized by requiring the substrate theory
$T_{\mathsf{S}}$ to be such that: 

$$\text{(\textbf{C1})} \quad\quad\quad T_{\mathsf{S}}\models \forall x.\bigvee_{c\in C} x=c\wedge \bigwedge_{c_1\in C,c_2\in C\setminus\{c_1\}} c_1\neq c_2$$
where $C$ is a finite set of constants of sort $\principal$. 
This imposes that there are exactly $|C|$ principals. 

The second condition
concerns the form of the policies: $(\textbf{C2})$ {for each CLP rule in
$\mathit{Po}$, all the variables in its body but not in its head range
over the set $C$ of principals.} 
For the fix-point computation required to solve an instance of the
reachability problem, variables not occurring in the head of a CLP
rule but only in its body must be eliminated by a suitable (quantifier
elimination) procedure (see, e.g.,~\cite{constraintdatalog}).
Assuming that such variables range only over the set $C$ of
principals---cf.~condition (\textbf{C1})---it is possible to replace
each one of them with the constants in $C$ and take disjunction.

The third and fourth conditions state  respectively that:  $(\textbf{C3})$ {$T_{\mathsf{S}}$ must be closed under sub-structures (see Section \ref{app:infprelim} in Appendix) and $(\textbf{C4})$
$T_{\mathsf{S}}$ must be {locally finite} (see Section \ref{app:infprelim} in Appendix). 
Examples of effectively locally finite
theories are the theory of an enumerated data-type or the theory of
linear orders (cf., e.g., \cite{jsc-ftp09}) for more details).
The last two (more technical) conditions allow us to reduce the
satisfiability of a formula containing universal quantifiers (namely,
those in the CLP rules) to the satisfiability of ground formulae by
instantiating variables with finitely many (representative) terms.
This implies the decidability of the satisfiability modulo
$T_{\mathsf{S}}$ of (\ref{eq:bmc}) (in the definition of reachability
problem) provided that it is decidable to check the satisfiability
modulo $T_{\mathsf{S}}$ of ground formulae.
\begin{theorem}
\label{thm3}
  Let $\mathit{Po}$ be a finite set of policies, $G$ a query, and
  $H_0,\ldots, H_{n-1}$ a sequence of non-mechanizable facts ($n\geq
  1$).  If (\textbf{C1}), (\textbf{C2}), (\textbf{C3}), and
  (\textbf{C4}) are satisfied and the satisfiability modulo
  $T_{\mathsf{S}}$ of ground formulae is decidable, then the instance
  of the bounded reachability analysis problem (with bound $n$,
  sequence $H_0, \ldots, H_{n-1}$ of non-mechanizable facts, and query
  $G$) is decidable.
\end{theorem}
The proof of this result uses previous work~\cite{jsc-ftp09} (see Section \ref{proof} in the Appendix)
and yields the correctness of the automated analysis technique in
Fig.~\ref{fig:automated-analysis}.  

This is only a first step towards
the design of a usable automated technique.  In fact, at each
iteration of the procedure, the solution of the bounded reachability
problem (at step 1(c)) requires to compute a fix-point and to check
the satisfiability modulo the substrate theory.  Such activities can
be computationally quite expensive and any means of reducing their
number is obviously highly desirable.

\section{A Reduction Technique}  
\label{sec:verification}
The main drawback of the procedure in Fig.~\ref{fig:automated-analysis}
is step 1 that forces the enumeration of all sequences in $\Sigma$.
Unfortunately, $\Sigma$ can be very large, e.g., there are 12 execution paths of \CRO{}
that are compatible with the MSC in Fig.~\ref{fig:msc-cro}.
To overcome this problem, in the rest of
this section, we design a reduction technique that allows for the
parallel execution of a group of ``independent'' exchanges of messages
so that several sequences of $\Sigma$ can be considered at the same time
in one iteration of step 1 in the algorithm of
Fig.~\ref{fig:automated-analysis}. In this way, the number of fix-point
computations and satisfiability checks may be significantly reduced. The
key to this refinement is a compact representation of the set $\Sigma$
based on an adaptation of Lamport's \emph{happened before} relation
$\leadsto$ \cite{lamport-comm}. There are many possible choices to
describe $\Sigma$, ranging from MSCs (as done in the previous section)
to BPEL workflows for web services augmented with access control
information~\cite{pacibertino}. We have chosen $\leadsto$ as a starting
point because it is at the same time simple and general, and simplifies
the design of our reduction technique.

\begin{figure}[t]
\centering
\begin{minipage}{.9\textwidth}
\underline{\textsf{Input}}: a substrate theory $T_{\mathsf{S}}$, a set
$\mathit{Po}$ of policies, and a scenario = (a finite set $C$ of
principals, a set $\Sigma$ of sequences of state-change rules of
finite length, a query $G$)\\
\underline{\textsf{Output}}: $G$ is reachable/unreachable\\
\underline{\textsf{Assumptions}}:  (\textbf{C1}), (\textbf{C2}),
(\textbf{C3}), and  (\textbf{C4}) are satisfied
\begin{enumerate}
\item For each sequence $\sigma\in \Sigma$:
\begin{enumerate}
  \item Determine the sequence $H_0, ..., H_{|\sigma|}$ of
    non-mechanizable facts that enables the corresponding sequence of
    instances of the state-change rule (\ref{eq:send-action}).
  \item Build an instance of the bounded reachability problem with
    bound $|\sigma|$, the non-mechanizable facts of the previous
    step, and the given query $G$.
  \item Try to solve the instance of the bounded reachability problem
    built at previous step.
  \item If one of the instances at the previous step turns out to be
    solvable, return that the query $G$ is reachable.
\end{enumerate}
\item Return that the query $G$ is unreachable (if step 1(d) is never executed).
\end{enumerate}
\end{minipage}
\caption{\label{fig:automated-analysis}Automated Analysis of Scenario-based Specifications (interleaving semantics)}
\end{figure}

\subsection{The Causality Relation} 
$\leadsto$ is a means of ordering a set
$L$ of events based on the potential causal relationship of pairs of
events in a concurrent system. Formally, $\leadsto$ is a partial order
on $L$, i.e., it is irreflexive, ($l\not\leadsto l$ for $l\in L$),
transitive  
(if $l_1\leadsto l_2$ and $l_2\leadsto l_3$ then $l_1\leadsto
l_3$ for $l_1, l_2, l_3\in L$), and 
anti-symmetric (if $l_1\leadsto l_2$ then $l_2\not \leadsto l_1$ for $l_1, l_2\in L$). 

Two distinct events
$l_1$ and $l_2$ are \emph{concurrent} if $l_1\not\leadsto l_2$ and
$l_2\not\leadsto l_1$ (i.e., they cannot causally affects each other). 
In the usual interleaving semantics, the set of possible executions can be seen as the set of all
linear orders that extend $\leadsto$. Formally, $\leadsto_t$ is a linear
extension of $\leadsto$ if $\leadsto_t$ is a total order 
(i.e., a partial order that is also total, for every $l_1,l_2\in L$ we have that
$l_1\leadsto_t l_2$ or $l_2\leadsto_t l_1$) 
that preserves $\leadsto$ (i.e., for every $l_1,l_2\in L$, if $l_1\leadsto l_2$, then $l_1\leadsto_t l_2$). 

For example, if $L=\{l_1,l_2\}$, $l_1$ and $l_2$
are \emph{concurrent}, then both $l_1\leadsto_t l_2$ and $l_2\leadsto_t
l_1$ are possible linear extensions since $l_1$ and $l_2$ cannot
causally affects each other. Enumerating all the elements of the set
$E(\leadsto)$ of linear extensions of the partial order $\leadsto$ can
be done in $O(|E(\leadsto)|)$ constant amortized
time~\cite{pruesse-ruskey} and computing $|E(\leadsto)|$ (i.e., counting
the number of linear extensions of $\leadsto$) is
\#P-complete~\cite{brightwell-winkler}.  

In our framework, $L$ is the set of instances
of the state-change rule (\ref{eq:send-action}) considered in a
scenario-based specification. 
Thus, the relation $\leadsto$ must be specialized so that
the following two constraints must hold: 
\begin{description}
\item[(\textbf{COMP1})] the
enabledness of concurrent (according to $\leadsto$) instances of
(\ref{eq:send-action}) must be preserved---i.e., two such instances
should not to causally affect each other by enabling (disabling) a
disabled (enabled, respectively) state-change rule,
\item[(\textbf{COMP2})] any execution of the concurrent events in a (finite)
set $L$, each of which causally affects another event $l$ not in $L$,
results in a state in which $l$ is enabled.
\end{description}
 
These requirements are
formalized as follows.
\begin{description}
\item[(COMP1)] if $l_2$ is enabled (disabled) in state $S$ then it is
  still enabled (disabled, respectively) in state $S'$ for
  $S\rightarrow_{l_1} S'$ and the same must hold when swapping $l_1$
  with $l_2$.
\item[(COMP2)] $Pre(l')\subseteq L$ is such that $l\leadsto^* l'$ for
  each $l\in Pre(l')$ 
  and there is no $l''\in L\setminus Pre(l')$ such that
  $l''\leadsto^* l'$, then $l'$ is enabled in $S'$ where
  $S\rightarrow_{l_1} \cdots \rightarrow_{l_k} S'$ and $Pre(l')=\{l_1,
  ..., l_k\}$.
\end{description}
(\textbf{COMP1}) implies that the execution of either $l_1$ or $l_2$
followed by $l_2$ or $l_1$, respectively, will produce two identical
states provided that the two executions start from the same initial
state.  
(\textbf{COMP2}) says that once the action of sending a
message is enabled, it persists to be so; this is related to the fact
that of (\ref{eq:send-action}) can only add messages to $\msg$.
Although (\textbf{COMP2}) seems to be restrictive at first sight, it
is adequate for checking reachability (safety) properties as we do in
this paper.

\begin{definition}[Causality Relation]
A partial order relation $\leadsto$ on a finite set $L$ of instances
of (\ref{eq:send-action}) that satisfies (\textbf{COMP1}) and
(\textbf{COMP2}) is called a causality relation.  
\end{definition}

The tuple
$(C, L, \leadsto, G)$ identifies a scenario ($C$, $\Sigma$, $G$) for
$C$ a finite set of principals, $G$ a ground query, and $\Sigma$ is
the set of sequences obtained by enumerating all the linear extensions
of $\leadsto$ on $L$.  
Since any linear extension of $\leadsto$ is of finite length (as
$\leadsto$ is acyclic), we will also call $(C, L, \leadsto, G)$ a
scenario.

We observe that when the state $S$ is given, it is possible to show
that both (\textbf{COMP1}) and (\textbf{COMP2}) are decidable (the
proof is similar to that of Theorem~\ref{thm3}).  In practice, it is
not difficult to argue that (\textbf{COMP1}) and (\textbf{COMP2}) hold
for a given scenario.  

\begin{example}
To illustrate this, we reconsider the scenario
informally specified in Fig.~\ref{fig:msc-cro} for the \CRO
\begin{figure}[t] 
  \begin{tabular}{ccc}
    \begin{minipage}{2.9cm}
$C := \{\Ed,\Helen,$ \\
\quad\quad $\theCA, \CRep\}$\\
\ \\
$L := \{\mathit{SEC},\mathit{SEC}_2$\\
$\mathit{SHC}, \mathit{SHC}_2,
        \mathit{SPC}
      \}$
\\

$\leadsto$ is the smallest partial order s.t.: \

     \begin{center}
      - $\mathit{SEC}\leadsto \mathit{SEC}_2$ \\
      - $\mathit{SHC}\leadsto \mathit{SHC}_2$\\
      - $\mathit{SPC}\leadsto \mathit{SPC}_2$
     \end{center}
    \end{minipage} 
    & \quad &
    \begin{minipage}{.75\textwidth}
      {\footnotesize
      \begin{tabular}{|r|l|}
        \hline
        $(SEC)$ & $\know(\theCA, \atoi(\Ed,\isemployee)) \Rightarrow
                   \oplus \msg(\theCA,\said(\atoi(\Ed,\isemployee)),\Ed)$ 
        \\ \hline
        $(SHC)$ & $\know(\theCA, \atoi(\Helen,\ishead)) \Rightarrow$ \\
                & $\,\quad\quad\quad\quad\quad\quad\quad\quad\quad \oplus \msg(\theCA,\said(\atoi(\Helen,\ishead)),\Ed)$
        \\ \hline
        $(SPC)$ & $\know(\Helen, \atoi(\Ed,\canstoredoc)) \Rightarrow$ \\
                & $\,\,\quad\quad\quad\quad\quad\quad\quad\quad    \oplus \msg(\Helen,\said(\atoi(\Ed,\canstoredoc)),\Ed)$
        \\ \hline
        $(SEC_2)$ & $\know(\Ed, \stoi(\theCA, \said(\atoi(\Ed,\isemployee))))
        \Rightarrow$ \\
                  & $\quad\quad\quad\quad\oplus \msg(\Ed,\said(\stoi(\theCA, \said(\atoi(\Ed,\isemployee)))),\CRep)$
        \\ \hline
        $(SHC_2)$ & $\know(\Ed, \stoi(\theCA, \said(\atoi(\Helen,\ishead)))) \Rightarrow$ \\
                  & $\,\,\quad\quad \oplus \msg(\Ed,\said(\stoi(\theCA, \said(\atoi(\Helen, \ishead)))),\CRep)$
        \\ \hline
        $(SPC_2)$ & $\know(\Ed, \stoi(\Helen, \said(\atoi(\Ed,\canstoredoc)))) \Rightarrow$ \\
        & $\,\,\,\quad \oplus \msg(\Ed,\said(\stoi(\Helen, \said(\atoi(\Ed, \canstoredoc)))),\CRep)$
        \\ \hline
        \end{tabular}}
    \end{minipage}
    \end{tabular}
  \caption{\label{fig:msccerts}Formalization of the \CRO{} scenario in Fig.~\ref{fig:msc-cro}}
\end{figure}
and recast it in the formal framework developed above, as shown in
Fig.~\ref{fig:msccerts}.  

There is an obvious correspondence between
the entries of the tables in the two figures. 
The message $m_1$ is the result of executing $SEC$, $m_2$ of $SHC$, $m_3$ of $SPC$, $m_4$
of $SEC_2$, $m_5$ of $SHC_2$, and $m_6$ of $SPC_2$.  There is also a
correspondence between the MSC in Fig.~\ref{fig:msc-cro} and the
causality relation $\leadsto$ in Fig.~\ref{fig:msccerts}.  

Now, we
show that the requirement (\textbf{COMP1}) holds for each pair $(l_1,l_2)$ of
concurrent rule instances in $L$ as follows.  For $(SEC)$ and $(SHC)$,
we have that if the latter is enabled (disabled) before the execution
of the first $(SHC)$, it remains enabled (disabled, respectively)
after its execution; the vice versa also holds.  Similar observations
hold also for the remaining pairs of concurrent events in $L$.  

Then, we show that the requirement (\textbf{COMP2}) holds for the events $(SEC)$ and
$(SEC_2)$ that are such that $(SEC)\leadsto (SEC_2)$.  
When $(SEC)$ is
executed, $(SEC_2)$ becomes enabled since the fact
$\msg(\theCA,\said(\atoi(\Ed,\isemployee)), \Ed)$ holds as the result
of executing $(SEC)$: by the CLP rule (\ref{eq:receive-action})
it is possible to derive $\know(\Ed,\stoi(\theCA, \said(\atoi(\Ed,\isemployee))))$
that is precisely the enabling condition of $(SEC_2)$.  Similar observations hold for
$(SHC)$ and $(SHC_2)$ as well as $(SPC)$ and $(SPC_2)$.  Intuitively,
$\leadsto$ formalizes the obvious remark that, before $\Ed$ can forward
a certificate to $\CRep$ (about his role, $\Helen$'s role, or the
permission to store documents), he must have preliminarily received it
regardless of the order in which he has received the certificates from
$\theCA$ and $\Helen$.\qed
\end{example}

\subsection{A reduction technique based on causality relations.} 

So far, we have shown that a causality relation can be exploited to compactly
specify a scenario. Here, we show how it can be used to dramatically
reduce the number of fix-point computation and satisfiability checks
required by the analysis technique in
Fig.~\ref{fig:automated-analysis} while preserving its
completeness. The key idea is the following. 

Since pairs of concurrent
rule instances cannot causally affect each other, it is possible to
execute them in \emph{parallel}, i.e.\ adopting a partial order
semantics. In fact, any linearization of the parallel execution, in
the usual interleaving semantics, will yield the same final state
obtained from the parallel execution. This has two advantages. 

First,
a single parallel execution of concurrent events correspond to a
(possibly large) set of linear executions. Second, the length of the
parallel execution is shorter than those of the associated linear
executions.  The number of fix-point computations and satisfiability
checks needed to solve a bounded reachability problem can be reduced
depending on the degree of independence of the rule instances in the
scenario. The price to pay is a modification of the definition of
reachability problem (cf.~the end of Section~\ref{sec:DKAL-light}) to
adopt a partial order semantics. We explain in more detail these ideas
below.

Let $\mathit{Po}$ be a (finite) set of policies and $(C, L, \leadsto,
G)$ a scenario, where $C$ is a finite set of principals, $L$ is a
finite set of rule instances of (\ref{eq:send-action}), $\leadsto$ is
a causality relation, and $G$ a query.

\begin{definition}[Reachability problem with partial-order semantics compatible
  with the causality relation $\leadsto$]
  \label{def:reachprobswithpartorder}
An instance of this problem 
amounts to establishing whether there exist an integer $n\geq 0$ and (ground) constraint facts
$H_0(\knowzero_0), ...,$ $H_{n-1}(\knowzero_{n-1})$ s.t.~$\mathcal{R}_n\cup \{ G(\know_n) \} ~ \text{is satisfiable modulo}
~ T_{\mathsf{S}}$,
where $\mathcal{R}_0 :=
\mathit{lfp}(\{H_0(\knowzero_0)\} \cup \mathit{Po}(\know_0))$,
$\mathcal{R}_{i+1} := \mathit{lfp}(\mathit{R}_{i+1} \cup
\{H_{i+1}(\knowzero_{i+1})\} \cup \mathit{Po}(\know_{i+1}))$, and
  $\mathcal{R}_i \rightarrow_{l_1} \cdots \rightarrow_{l_k} \mathit{R}_{i+1}$
for $l_1, ..., l_k\in L$ such that any pair $(l_a,l_b)$ is of
concurrent events ($a,b=1, ..., k$ and $a\neq b$).
\end{definition}

Definition \ref{def:reachprobswithpartorder} is
almost identical to to the Definition \ref{def:instofreachprob}. The main
difference is in allowing the execution of a sequence $l_1, ..., l_k$
of exchange of messages provided that these are pairwise concurrent
with respect to the causality relation.  Intuitively, we cumulate the
effect of executing the instances $l_1, ..., l_k$ of
(\ref{eq:send-action}) in a single step so that each principal can
derive more knowledge from the exchange of several messages than the
exchange of just one message as it was the case with the definition of
reachability problem in Section~\ref{sec:DKAL-light}.

With this new definition of reachability problem, we propose a
refinement in Fig.~\ref{fig:automated-analysis-refinement} of the
analysis technique in Fig.~\ref{fig:automated-analysis}.
\begin{figure}[t]
\centering
\begin{minipage}{\textwidth}
\underline{\textsf{Input}}: a substrate theory $T_{\mathsf{S}}$, a set
$\mathit{Po}$ of policies, and a scenario
 $(C, L, \leadsto, G)$= (a finite set $C$ of
principals, a finite set $L$ of instances of (\ref{eq:send-action}), a causality relation $\leadsto$, a query $G$)\\
\underline{\textsf{Output}}: $G$ is reachable/unreachable\\
\underline{\textsf{Assumptions}}:   (\textbf{C1}), (\textbf{C2}),
(\textbf{C3}), (\textbf{C4}), (\textbf{COMP1}), and (\textbf{COMP2}) are satisfied
\begin{enumerate}
\item Let $CG(\leadsto)$ be the causality graph associated to
  $\leadsto$.
\item Compute the set $P_0$ of nodes in $CG(\leadsto)$ with no
  incoming edges and $L_{P_0}$ be the set of rule instances of
  (\ref{eq:send-action}) in $L$ labeling the nodes in $P_0$.
\item Determine the set $H_0$ of non-mechanizable facts that enables
  all the rule instances in $L_{P_0}$.
\item Set $j=0$ and while the set of nodes in $CG(\leadsto)$ is
  non-empty do
  \begin{enumerate}
    \item Delete from $CG(\leadsto)$ all the nodes in $P_j$ and the
      edges whose sources are in $P_j$ and increment $j$ by $1$.
    \item Compute the set $P_j$ of nodes in $CG(\leadsto)$ with no
      incoming edges and $L_{P_j}\subseteq L$ be the set of rule
      instances labeling the nodes in $P_j$.  
  \end{enumerate}
\item Build an instance of the bounded reachability problem with
  partial order semantics compatible with the causality relation
  $\leadsto$ with bound $j$, sequence $H_0, H_1, ..., H_{j}$ of
  non-mechanizable facts where $H_i:=\mathit{true}$ for $i=1, ...,
  j$, and the input query $G$.  At each step $i$ of the bounded
  reachability problem, the rule instances in $L_{P_j}$ must be used
  for parallel execution.
\item If the instance of the bounded reachability problem is solvable,
  then return that the query $G$ is reachable; otherwise, return that
  $G$ is unreachable.
\end{enumerate}
\end{minipage}
\caption{\label{fig:automated-analysis-refinement}Automated Analysis of Scenario-based Specifications (partial order semantics)}
\end{figure}
The main differences between the two techniques are the following.  In
input, the scenario is given by using the notion of causality relation
in order to exploit the new definition of reachability problem with
the partial order semantics.  Then, instead of considering all the
possible linear extensions of $\leadsto$ (as in
Fig.~\ref{fig:automated-analysis}), sets of pairwise concurrent events
for parallel execution are computed by using the causality relation.
The idea is to use the Hasse diagram $CG(\leadsto)$, called the
\emph{causality graph} in the following, associated to $\leadsto$,
i.e.\ the transitive reduction of the relation $\leadsto$ seen as an
oriented graph.  The crucial observation is that concurrent events can
be identified by looking at those nodes that are not connected by a
path in the causality graph.  Formally, we need the following notion.
An element $l$ is \emph{minimal} in $L$ with respect to $\leadsto$ iff
there is no element $l'\in L$ such that $l'\leadsto l$.  Since $L$ is
finite, minimal elements of $\leadsto$ must exist (this is a basic
property of partial orders over finite sets).  In step 2, the rule
instances labeling the minimal elements in $L$ with respect to
$\leadsto$, that correspond to nodes with no incoming edges in the
causality graph, are the only that require non-mechanizable facts for
them to be enabled.  In fact, all the rule instances labeling
non-minimal elements with respect to the causality are enabled by the
execution of one or more rule instances that label ancestor nodes in
the causality graph, because of (\textbf{COMP2}).  This is why we
compute $H_0$ in step 3 while all the other sets of non-mechanizable
facts are vacuosly set to $\mathit{true}$ in step 5.  The rule
instances in $L_{P_0}$ labeling the nodes in $P_0$ are concurrent
because of (\textbf{COMP1}).  In step 4, we exploit this observation
to compute the other set of concurrent rule instances that can be
executed in parallel by modifying the causality graph: the nodes and
the edges whose sources are in $P_0$ are deleted from $CG(\leadsto)$
so that a the set $P_1$ of nodes with no incoming edges can be
identified.  The rule instances in $L_{P_1}$ labeling the nodes in
$P_1$ are the new concurrent events that can be executed in parallel
and so on.  The procedure eventually terminates when no more nodes are
left in the causality graph.  Then, in step 5, the new definition of
bounded reachability problem compatible with the causality relation
$\leadsto$ can be exploited by using the sets $L_{P_0}, ..., L_{P_j}$
of rules instances to be executed in parallel.  If the instance is
solvable then the query $G$ is reachable, otherwise it is unreachable.
The correctness of the refined analysis in
Fig.~\ref{fig:automated-analysis} stems from the fact that, by
definition of causality relation, there exists an execution in the
interleaving semantics for the concurrent events executed in
parallel---because of (\textbf{COMP1})---and the execution of rule
instances that must happen before (with respect to $\leadsto$), enable
the execution of those that happen afterwards---according to
(\textbf{COMP2}).

We briefly illustrate how the refined version of the automated
analysis works on the scenario in Fig.~\ref{fig:msc-cro} of the
\CRO{}.  According to the causality relation in
Fig.~\ref{fig:msccerts}, $(SEC), (SHC)$, and $(SPC)$ are minimal
elements of $\leadsto$ (step 2).  Thus, the non-mechanizable fact
$H_0$ enabling their execution is the conjunction of the following
three facts: $\knowzero(\theCA, \atoi(\Ed,\isemployee))$,
$\knowzero(\theCA, \atoi(\Helen,\ishead))$, and $\knowzero(\Helen,
\atoi(\Ed,\canstoredoc))$.  Deleting the nodes labeled by $(SEC),
(SHC)$, and $(SPC)$ with the corresponding edges in the causality
graph leaves us with a graph containing three isolated nodes labeled
by $(SEC_2), (SHC_2)$, and $(SPC_2)$ that can be executed in parallel.
As a consequence, the bound of the reachability problem is $2$ in
which, initially, the parallel execution of $(SEC), (SHC)$, and
$(SPC)$ is enabled because of the non-mechanizable facts in $H_0$
while the parallel execution of $(SEC_2), (SHC_2)$, and $(SPC_2)$ is
enabled, in the following step, because of the three new certificates
available in the net.  Even in this simple example, the savings of the
reduction technique are important: the two-step parallel execution
corresponds to 
$6$ interleavings executions that must be considered
when using the technique in Fig.~\ref{fig:automated-analysis}.

We have implemented a prototype of the procedure above in
WSSMT~\cite{BCRVZ} that uses the SMT solver Z3~\cite{z3} for fix-point
computation and SMT solving.  The time taken to analyze the scenario
in Fig.~\ref{fig:msccerts} with this prototype is negligible; larger
examples are discussed in~\cite{BCRVZ}.

\section{Discussion}
\label{sec:conclusions}
We presented an automated technique to analyze scenario-based
specifications of access control policies in open and distributed
systems that takes into account human activities.  It uses an instance
of CLP to express policies and trust relationships, and reduces the
analysis problem to fix-point computations and satisfiability checks.
The first contribution is the decidability of the analysis of
scenario-based specifications of ACSs.  The second contribution is a
reduction technique that allows us to make the decidability result
useful in practice.\\



There are three main lines of research that are related to our work. 
First, several logic-based frameworks
(e.g.,~\cite{lietal2005,GurevichNeeman-dkal,becker,hurlin,asiaccs11,jsc-ftp09})
have been proposed to specify and analyze authorization policies with
conditions depending on the environment of the system in which they
are enforced.  In principle, it is possible to consider the conditions
depending on the execution of human activities as part of the
environment and then re-use the available specification and analysis
techniques.  The problem in doing this is that the conditions for the
execution of human activities are not explicitly modeled in the
system so that their applicability is unconstrained. This results in
a dramatic increase of the search space that makes the application of
the available technique difficult, if possible at all.  We avoid this
state-explosion problem by considering scenario-based specifications
that allow one to focus on a small sub-set of the possible sequences of
events, as explained in Section~\ref{sec:verification}.  It would be
interesting to adapt the abduction techniques
in~\cite{becker-abd,hurlin} to identify which non-mechanizable facts
need to be generated for the executability of complex scenarios in
which condition (\textbf{COMP2}), about the ``monotonicity'' of the
events (Section~\ref{sec:verification}), does not hold.

The second line of research is related to workflow analysis in
presence of authorization policies, e.g.,~\cite{pacibertino,crampton}.
On the one hand, such works specify the workflow as a partial ordering
on tasks that is similar to the causality relation introduced here.
On the other hand, these works abstract away the data-flow so that
there is no need to specify compatibility conditions on the causality
relation (cf.~(\textbf{COMP1}) and (\textbf{COMP2}) in
Section~\ref{sec:verification}) as we do here because of the modelling
of the exchange of messages among principals.  Another difference is
that the specification of authorization policies is reduced to a
minimum in~\cite{pacibertino,crampton} so as to simplify the study of
the completion problem, i.e., whether there exists at least one
assignment of users to tasks that allow for the execution of the whole
workflow.  Instead, we focus on reachability problems and we model,
besides authorization policies, also trust relationship among
principals.  It would be interesting to study the decidability of the
completion problem also in our richer framework.  

The third line of research concerns the development of (semi) formal techniques
for the analysis of human interventions.
In \cite{diaper90,suokas85} the authors aim to determine how a task is executed by
humans and what special factors are involved to accomplish the goal
the task is supposed to achieve. This line of work is based on informal methods 
to identify and analyze human actions in contrast to our framework that is based
on a logical formalism. 
In \cite{shin06} the authors use graphs and deterministic finite state automata
to model and analyze human behaviors in critical systems.
Although we share the proposed formal approach with them, our framework differs for the capability to analyze systems influenced by non predictable human activities, 
in contrast with those predefined for industrial material-handling processes.
Interesting works in modeling and reasoning about human operators 
are, e.g., \cite{yasmeen11,gunter09}, where the analysis is based on concurrent game structures, a formalism similar to the $\acs$ we used in Section~\ref{sec:DKAL-light}.
The accurate verification analysis and the decidability result we presented in this paper 
are the major difference that distinguishes our work from their.

\bibliographystyle{abbrv}
\bibliography{biblioext}

\begin{thebibliography}{10}

\bibitem{asiaccs11}
F.~Alberti, A.~Armando, and S.~Ranise.
\newblock {Efficient Symbolic Automated Analysis of Administrative Role Based
  Access Control Policies}.
\newblock In {\em Proc.~6th ASIACCS}. ACM Press, 2011.

\bibitem{avantssar}
AVANTSSAR.
\newblock {Deliverable 5.1: Problem cases and their trust and security
  requirements}.
\newblock Available at http://www.avantssar.eu, 2008.

\bibitem{BCRVZ}
M.~Barletta, A.~Calvi, S.~Ranise, L.~Vigan{\`o}, and L.~Zanetti.
\newblock {WSSMT: Towards the Automated Analysis of Security-Sensitive Services
  and Applications}.
\newblock In {\em Proc.~12th SYNASC}, pages 417--424. IEEE Computer Society,
  2010.

\bibitem{BRV-TR09}
M.~Barletta, S.~Ranise, and L.~Vigan\`o.
\newblock {Verifying the Interplay of Authorization Policies and Workflow in
  Service-Oriented Architectures}.
\newblock In {\em Proc.~SecureCom 2009}, pages 289--299. IEEE CS Press, 2009.
\newblock Full version at http://arxiv.org/abs/0906.4570.

\bibitem{BRV-TR12}
M.~Barletta, S.~Ranise, and L.~Vigan\`o.
\newblock {Automated Analysis of Scenario-based Specifications of Distributed
  Access Control Policies with Non-Mechanizable Activities (Extended Version)}.
\newblock Technical report, Dip. di Inf., Univ. di Verona, Italy, 2012.

\bibitem{becker-abd}
M.~Y. Becker and S.~Nanz.
\newblock The role of abduction in declarative authorization policies.
\newblock In {\em Proc. of the 10th int. conf. on Practical aspects of
  declarative languages (PADL'08)}, pages 84--99. Springer-Verlag, 2008.

\bibitem{becker}
M.~Y. Becker and S.~Nanz.
\newblock A logic for state-modifying authorization policies.
\newblock {\em ACM Trans. Inf. Syst. Secur.}, 13(3), 2010.

\bibitem{pacibertino}
E.~Bertino, J.~Crampton, and F.~Paci.
\newblock {Access Control and Authorization Constraints for WS-BPEL}.
\newblock In {\em Proceedings of ICWS'06}, pages 275--284. IEEE Computer
  Society Press, 2006.

\bibitem{brightwell-winkler}
G.~Brightwell and P.~Winkler.
\newblock {Counting linear extensions is \#P-complete}.
\newblock In {\em Proc. of the 23rd ACM Symp. on Theory of Comp.}, STOC '91,
  pages 175--181. ACM, 1991.

\bibitem{diaper90}
D.~Diaper.
\newblock {\em Task analysis for human-computer interaction}.
\newblock Prentice Hall, 1990.

\bibitem{dkalimpl}
{Dkal Code Plex}.
\newblock \url{http://dkal.codeplex.com/}.

\bibitem{enderton}
H.~B. Enderton.
\newblock {\em A Mathematical Introduction to Logic}.
\newblock Academic Press, 1972.

\bibitem{gunter09}
E.~Gunter, A.~Yasmeen, C.~Gunter, and A.~Nguyen.
\newblock Specifying and analyzing workflows for automated identification and
  data capture.
\newblock In {\em In Proc. of HICSS'09}, pages 1--11. IEEE, 2009.

\bibitem{GurevichNeeman-dkal}
Y.~Gurevich and I.~Neeman.
\newblock Dkal: Distributed-knowledge authorization language.
\newblock In {\em Proceedings of CSF}, pages 149--162, 2008.

\bibitem{hurlin}
C.~Hurlin and H.~Kirchner.
\newblock Semi-automatic synthesis of security policies by invariant-guided
  abduction.
\newblock In {\em Proc. of the 7th Int conf. on Formal aspects of security and
  trust (FAST'10)}, pages 157--175. Springer-Verlag, 2011.

\bibitem{lamport-comm}
L.~Lamport and F.~B. Schneider.
\newblock Pretending atomicity.
\newblock Technical report, In Research Report 44, Digital Equipment
  Corporation Systems Research, 1989.

\bibitem{constraintdatalog}
N.~Li and J.~Mitchell.
\newblock {Datalog with constraints: a foundation for trust management
  langauges}.
\newblock In {\em PADL'03}, LNCS 2562, pages 58--73. Springer, 2003.

\bibitem{beyond-proof-compliance}
N.~Li, J.~C. Mitchell, and W.~H. Winsborough.
\newblock Beyond proof-of-compliance: security analysis in trust management.
\newblock {\em J. ACM}, 52:474--514, May 2005.

\bibitem{lietal2005}
N.~Li and M.~V. Tripunitara.
\newblock Security analysis in role-based access control.
\newblock {\em ACM Trans. Inf. Syst. Secur.}, 9:391--420, November 2006.

\bibitem{MauSch96}
U.~M. Maurer and P.~E. Schmid.
\newblock {A Calculus for Security Bootstrapping in Distributed Systems}.
\newblock {\em Journal of Computer Security}, 4(1):55--80, 1996.

\bibitem{msr}
{Microsoft Research}.
\newblock \url{http://research.microsoft.com/en-us/}.

\bibitem{pruesse-ruskey}
G.~Pruesse and F.~Ruskey.
\newblock {Generating Linear Extensions Fast}.
\newblock {\em SIAM J. Comp.}, 23(2):373--386, 1994.

\bibitem{ramsey}
F.~P. Ramsey.
\newblock {On a Problem of Formal Logic}.
\newblock {\em Proceedings of the London Mathematical Society},
  s2-30(1):264--286, 1930.

\bibitem{jsc-ftp09}
S.~Ranise.
\newblock {On the Verification of Security-Aware E-services}.
\newblock {\em Journal of Symbolic Computation, Special Issue dedicated to the
  FTP'09 workshop}, 2012.
\newblock To appear.

\bibitem{rudolph96}
E.~Rudolph, P.~Graubmann, and J.~Grabowski.
\newblock Tutorial on message sequence charts.
\newblock {\em Computer Networks and ISDN Systems}, 28(12):1629--1641, 1996.

\bibitem{shin06}
D.~Shin, R.~Wysk, and L.~Rothrock.
\newblock Formal model of human material-handling tasks for control of
  manufacturing systems.
\newblock {\em Systems, Man and Cybernetics, Part A: Systems and Humans, IEEE
  Transactions on}, 36(4):685--696, 2006.

\bibitem{suokas85}
J.~Suokas.
\newblock {\em On the reliability and validity of safety analysis}.
\newblock Technical Research Centre of Finland, 1985.

\bibitem{crampton}
K.~Tan, J.~Crampton, and C.~Gunter.
\newblock The consistency of task-based authorization constraints in workflow.
\newblock In {\em Computer Security Foundations Workshop, 2004. Proceedings.
  17th IEEE}, pages 155--169, 2004.

\bibitem{TinelliZarba}
C.~Tinelli and C.~G. Zarba.
\newblock Combining non-stably infinite theories.
\newblock {\em Journal of Automated Reasoning}, 34(3):209--238, 2005.

\bibitem{yasmeen11}
A.~Yasmeen and E.~Gunter.
\newblock Automated framework for formal operator task analysis.
\newblock In {\em Proc. of ISSTA '11}, 2011.

\bibitem{z3}
{Z3 Home Page}.
\newblock \url{http://research.microsoft.com/en-us/um/redmond/projects/z3/}.

\end{thebibliography}

\section{Appendix}
\label{appx}
\subsection{Formal preliminaries}
\label{app:infprelim}

We assume some familiarity with the syntactic and semantic notions of
first-order logic~\cite{enderton} and Constraint
Datalog~\cite{constraintdatalog}.

Let $\Sigma$ be a signature, i.e., a collection of function and predicate symbols with their arities. 
A \emph{$\Sigma(\underline{x})$-expression} is an expression (a term, an
atom, a literal, or a formula) built out of the symbols in $\Sigma$
where at most the variables in the sequence $\underline{x}$ may occur
free. 
We write $E(\underline{x})$ to emphasize that $E$ is a
$\Sigma(\underline{x})$-expression. 
For two disjoint sequences of
variables $\underline{x}$ and $\underline{y}$, we write
$\underline{x},\underline{y}$ to denote their concatenation. 

A $\Sigma$-structure $\mathcal{N}$ is a \emph{sub-structure} of a
$\Sigma$-structure $\mathcal{M}$ iff the domain of $\mathcal{N}$ is
contained in the domain of $\mathcal{M}$ and the interpretations of
the symbols of $\Sigma$ in $\mathcal{N}$ are restrictions of the
interpretations of these symbols in $\mathcal{M}$. A class
$\mathcal{CL}$ of $\Sigma$-structures is \emph{closed under
  sub-structures} iff for every structure $\mathcal{M}\in
\mathcal{CL}$, if $\mathcal{N}$ is a substructure of $\mathcal{M}$
then $\mathcal{N}\in \mathcal{CL}$. 

Let $\Sigma$ and $\Sigma'$ be two
signatures such that $\Sigma\subseteq \Sigma'$. If $\mathcal{M}$ is a
$\Sigma'$-structure, then $\mathcal{M}|_{\Sigma}$ is the \emph{reduct}
of $\mathcal{M}$ obtained from $\mathcal{M}$ by forgetting the
interpretations of the symbols in $\Sigma'\setminus \Sigma$.

A \emph{$\Sigma$-theory} $T$ is a set of $\Sigma$-formulae, called
\emph{axioms}. A $\Sigma$-theory $T$ identifies a class
$\mathit{Mod}(T)$ of $\Sigma$-structures that are models of all
formulae in $T$. For each theory $T$ considered in the paper, we
assume that $\mathit{Mod}(T)\neq \emptyset$ and we then say that $T$
is \emph{consistent}. 

A $\Sigma$-formula $\varphi(\underline{x})$ is
\emph{$T$-satisfiable} iff there exists a $\Sigma$-structure
$\mathcal{M}\in \mathit{Mod}(T)$, also called a \emph{$\Sigma$-model},
such that $\mathcal{M} \models \exists \underline{x}.\,
\varphi(\underline{x})$. The \emph{satisfiability modulo theory $T$
  problem}, in symbols, $\mathit{SMT}(T)$, consists of establishing
the $T$-satisfiability of any quantifier-free $\Sigma$-formula. 

A $\Sigma$-theory $T$ is \emph{locally finite} if $\Sigma$ is finite
and, for every set of constants $\underline{a}$, there exist finitely
many ground terms $t_1, ..., t_{k_{\underline{a}}}$, called
\emph{representatives}, such that for every ground $(\Sigma\cup
\underline{a})$-term $u$, we have $T\models u = t_i$ for some $i$. If
the representatives are effectively computable from $\underline{a}$
and $t_i$ is computable from $u$, then $T$ is \emph{effectively}
locally finite. 
A theory $T$ admits \emph{quantifier elimination} if for an arbitrary formula $\varphi(\underline{x})$,
possibly containing quantifiers, one can compute a $T$-equivalent
quantifier-free formula $\varphi'(\underline{x})$.

A formula of the \emph{Bernays-Sch{\"o}nfinkel-Ramsey} (BSR) class is of
the form $\exists \underline{x}.\, \forall \underline{y}.\,
\varphi(\underline{x}, \underline{y})$, where $\underline{x},
\underline{y}$ are (disjoint) tuples of variables and $\varphi$ is a
quantifier-free formula built out of a signature containing only
predicate and constant symbols (i.e., no function symbol occurs in
$\varphi$). Formulae of the BSR class where $\underline{x}$ is empty are
called \emph{universal}, whereas when $\underline{y}$ is empty they are
called \emph{existential}. It is easy to show that any theory whose
axioms are universal BSR formulae is effectively locally finite. Satisfiability of
BSR formulae is well-known to be decidable~\cite{ramsey}.

Let $T$ be a $\Sigma$-theory and $\underline{R}$ a tuple of predicate
symbols not in $\Sigma$. A \emph{$\mathit{BSR}(T)$-formula} is a formula
of the form $\exists \underline{x}.\,\forall \underline{y}.\,
\varphi(\underline{x}, \underline{y})$, where $\varphi$ is a
quantifier-free $\Sigma^{\underline{R}}$-formula,
$\Sigma^{\underline{R}}:=\Sigma \cup \underline{R}$, and $\Sigma \cap
\underline{R}=\emptyset$. Universal and existential
$\mathit{BSR}(T)$-formulae are defined analogously to the corresponding
sub-classes of BSR formulae. A $\mathit{BSR}(T)$-formula $\psi$ is
\emph{$T$-satisfiable} iff there exists a
$\Sigma^{\underline{R}}$\,-structure $\mathcal{M}$ that satisfies $\psi$
and whose reduct $\mathcal{M}|_{\Sigma}$ is in $\mathit{Mod}(T)$.

\begin{theorem}[\cite{jsc-ftp09}]
  \label{th:dec-ext-BSR}
  Let $T$ be an effectively locally finite $\Sigma$-theory whose class of models is
  closed under sub-structures, the SMT($T$) problem be decidable, and
  $\underline{R}$ be a finite set of predicate symbols such that
  $\Sigma \cap \underline{R} = \emptyset$.  Then, the satisfiability
  of $\mathit{BSR}(T)$-formulae is decidable.
\end{theorem}

Let $T$ be a $\Sigma$-theory.  A \emph{constraint Datalog rule} is a
formula of the form
\begin{eqnarray*}
  \forall \underline{x},\underline{y}.\,
  \xi(\underline{x},\underline{y}) \wedge
  \bigwedge_{i=1}^n A_i(\underline{x},\underline{y}) 
  \rightarrow A_0(\underline{x})\,,
\end{eqnarray*}
also written as
\begin{eqnarray*}
  A_0(\underline{x})
  \ \leftarrow \
  A_1(\underline{x},\underline{y}) \wedge \cdots \wedge 
  A_n(\underline{x},\underline{y}) \wedge
  \xi(\underline{x},\underline{y})\,,
\end{eqnarray*}
where $A_i$ is an atom for $i=0,1,...,n$,
$\xi(\underline{x},\underline{y})$ is a quantifier-free
$\Sigma(\underline{x},\underline{y})$-formula, called the
\emph{constraint} of the rule, and $\underline{x},\underline{y}$ are
disjoint tuples of variables; when $n=0$, the constraint Datalog rule is
also called a \emph{constraint fact}.

The \emph{non-ground Herbrand base} of a set $\mathit{LP}$ of constraint
Datalog rules is the set of constraint facts modulo equality. The
\emph{non-ground immediate consequences operator}
$\mathcal{S}_{\mathit{LP}}$ is defined over a collection of constraint
facts $F$ as follows: $\mathcal{S}_{\mathit{LP}}(F)$ contains all the
constraint facts of the form $A_0(\underline{x}) \leftarrow
\xi(\underline{x},\underline{y})$ when $A_0(\underline{x}) \leftarrow
A_1(\underline{x},\underline{y}) \wedge \cdots \wedge
A_n(\underline{x},\underline{y}) \wedge
\xi(\underline{x},\underline{y})$ is in $\mathit{LP}$, $A_i\leftarrow
\xi_i'$ is in $F$ for $i=1, ..., n$, and $\xi$ is logically equivalent
(in $T$) to $\xi_1'\wedge \cdots \wedge \xi_n'$, where it is implicitly
assumed that the variables in the rule and those in the constraint facts
have been renamed so as to make them pairwise disjoint. It is possible
to show the existence of the least fix-point
$\mathit{lfp}({\mathit{LP}})$ of $\mathcal{S}_{\mathit{LP}}$, which may
be infinite.

It is sometimes possible to show that
$\mathit{lfp}(\mathcal{S}_{\mathit{LP}})$ is finite. Let $T$ admit
quantifier-elimination. Let $r_0(\underline{x}) \leftarrow
\bigwedge_{i=1}^n r_i(\underline{x})\wedge \xi_0(\underline{x})$ be a
constraint Datalog rule and $r_i(\underline{x}_{k_i}) \leftarrow
\xi_i(\underline{x}_{k_i})$ be a constraint fact for $\xi_i$ a
$\Sigma(\underline{x}_{k_i})$-quantifier-free formula, $k_i$ the arity
of $r_i$, and $i=1, ..., n$. A \emph{constraint rule application}
produces $m\geq 0$ facts of the form $r_0(\underline{x}) \leftarrow
\xi_j'(\underline{x})$ where $\xi_j'$ is a quantifier-free
$\Sigma(\underline{x})$-formula for $j=1, ..., m$ ($m\geq 0$) and
$\bigvee_{j=1}^m \xi_j'$ is equivalent (by the elimination of
quantifiers in $T$) to the formula
\begin{eqnarray*}
  \exists \underline{y}.\, (\bigwedge_{i=1}^n \xi_i(\underline{x}_{k_i})
    \wedge \xi_0(\underline{x})) ,
\end{eqnarray*}
where $\underline{y}$ is the tuple of variables occurring in the body
of the rule but not in the head.  The algorithm to compute the least
fix-point of a set of constraint Datalog rules is given in
Fig.~\ref{fig:fixpoint-comp}. 
The function $\mathsf{constrFP}$ terminates when all
derivable new facts are implied by previously derived facts so that
the least fix-point is reached.
\begin{theorem}[\cite{jsc-ftp09}]
  \label{coro:fixpoint-termination}
  Let $T$ be an effectively locally finite theory that admits
  elimination of quantifiers.  Then, $\mathsf{constrFP}$
  terminates returning a finite set of constraint facts.
\end{theorem}

\begin{figure}[t]
\footnotesize
  \centering
    \begin{minipage}{.3\textwidth}
      \begin{tabbing}
        foo \= foo \= foo \= foo \= foo \= foo \= foo \= foo \= foo \= foo \= \kill
        \textbf{function} $\mathsf{constrFP}(F,~ R)$ \\
        1 \> $\mathit{results}\leftarrow F$;  $\mathit{Changed}\leftarrow \mathit{true}$; \\ %
        2\> \textbf{while} $\mathit{Changed}$ \textbf{do}\\
        3\>\> $\mathit{Changed}\leftarrow \mathit{false}$\\
        4\>\> \textbf{foreach} $\mathit{rule}\in R$ \textbf{do} \\ 
        5\>\>\> \textbf{foreach} $\mathit{tuple}$ of constraint facts constructed from $\mathit{results}$  \textbf{do}\\
        6\>\>\>\> $\mathit{newres} \leftarrow $ constraint facts obtained by constraint rule  \\
        \>\>\>\> \hspace{2.05cm} application between $\mathit{rule}$ and $\mathit{tuple}$ \\
        7\>\>\>\> \textbf{foreach} $\mathit{fact}\in \mathit{newres}$ \textbf{do} \\ 
        8\>\>\>\> \textbf{if} $(\mathit{results} \not\models \mathit{fact})$ \textbf{then} $\mathit{results} \leftarrow \mathit{results}\cup \{\mathit{fact}\};$\\ 
        9 \>\>\>\> $\mathit{Changed} \leftarrow \mathit{true};$\\
        10\>\>\>\> \textbf{end} \\
        11\>\>\>\textbf{end}\\
        12\>\>\textbf{end}\\
        13\>\textbf{end}\\
        14\>\textbf{return} $\mathit{results}$
      \end{tabbing}
    \end{minipage}
  \caption{\label{fig:fixpoint-comp}Least fix-point computation of
    constraint Datalog rules (adapted from~\cite{constraintdatalog}):
    $F$ is a set of {constraint facts} and $R$ is a set of {constraint
      Datalog rules.}}
\end{figure}

\subsection{Proof of Theorem~\ref{thm3} }
\label{proof}
\begin{proof}
  By considering Definition \ref{acss} in Section \ref{sec:DKAL-light}, let $\psi_1, ..., \psi_{n-1}$ be a sequence (possibly containing repetitions) of elements in $\Psi$.  Since $\Psi$ is finite, there
  are finitely many sequences of elements of length $n$. 
  Thus, we can enumerate all of such sequences.  
  Since the value of $n$ is given together with the collection $\{H_0, ..., H_{n-1}\}$ of
  non-mechanizable facts, it is possible to compute the least fixed point of the set
  $\mathcal{R}_n$ of constraints in (\ref{eq:bmc}) by repeatedly
  invoking the function $\mathsf{constrFP}$ of
  Fig.~\ref{fig:fixpoint-comp} in Section \ref{app:infprelim} as follows.  Let
  $$\mathtt{R}_0:=\mathsf{constrFP}(\{I(\msg_0), H_0(\knowzero_0)\}, \widehat{\mathit{Po}}(\know_0))$$ 
be the set of constraint facts, generated from the initial state of the system, 
  where $\widehat{\mathit{Po}}$ has been obtained from $\mathit{Po}$
  by replacing each constraint Datalog rule $r$ with the set
  $\{r\sigma\}_{\sigma}$ where $\sigma$ ranges over the mappings that
  associate the variables occurring in the body of $r$ but not in the
  head of $r$ with the constants of sort $\principal$ in $C$, 
  satisfying the (\textbf{C1}) requirement.

By Theorem~\ref{coro:fixpoint-termination} in Section \ref{app:infprelim}, the invocation to
  $\mathsf{constrFP}$ terminates returning the finite set
  $\mathtt{R}_0$ of constraint facts.  
  In fact, both $I(\msg_0)$ and $H_0(\knowzero_0)$ are constraint facts,
  $\widehat{\mathit{Po}}(\know_0)$ is a set of constraint Datalog rules,
  and the substrate theory $T_{\mathsf{S}}$ satisfy the (\textbf{C4}) requirement.  
  Note that there is no need to eliminate quantifiers during a constraint rule application as all the variables
  occurring in the body of a constraint Datalog rule of
  $\widehat{\mathit{Po}}(\know_0)$ occur also in its head.  We are
  thus entitled to conclude that $\mathtt{R}_0$ and $\mathcal{R}_0$
  defined in \eqref{eq:bmc} are equal (modulo variable
  renaming).  Then, let
  \begin{align*}
    \mathtt{R}_1 = &~ \mathsf{constrFP}(\{H_1(\knowzero_1)\}, \mathtt{Eff}_1(\mathtt{R}_0,\psi_1(\msg_0,\msg_1)) \\
    & \cup \widehat{\mathit{Po}}(\know_1)), \text{ where, for }i=0,
  \end{align*}
 \begin{align*}
  & \mathtt{Eff}_i(F,\psi(\msg_i,\msg_{i+1}) = \{\msg_{i+1}(y,z,w) \leftarrow \mathit{Upd}_i(y,z,w)\sigma \mid \tilde{F}\cup \{G_i\sigma\} \text{ is} \\
   & \text{$T_{\mathsf{S}}$-satisfiable by considering $p_1, x_1, ..., p_m, x_m$} \text{as fresh constants}\}, \\
  & \tilde{F} := \{ A_0\leftrightarrow \xi ~|~ (A_0\leftarrow \xi)\in F \}\,,
  \end{align*}
  $\mathit{Upd}_i$ and $G_i$ are obtained from $\mathit{Upd}$ and $G$
  by replacing $\msg$ and $\knowzero$ with $\msg_i$ and $\knowzero_i$,
  respectively.  
  
  Note that $\mathtt{Eff}_i(F,\psi)$ is finite if $F$
  is so.  Also, the satisfiability of formulae over a signature
  extended with fresh constants is decidable when the satisfiability
  problem of the theory over its original signature is
  decidable~\cite{TinelliZarba}.\footnote{Note that we can extend
    the signatures with fresh (Skolem) constants since we consider all
    the classes of models of the substrate theory $T_{\mathsf{S}}$.
    The least fix-point semantics is used only when considering the
    constraint Datalog rules specifying the policies, not the
    substrate theory.}  Thus, since the
  $T_{\mathsf{S}}$-satisfiability is decidable by assumption, it is
  also decidable to check whether $\tilde{F}\cup \{G\sigma\}$ is
  $T_{\mathsf{S}}$-satisfiable. Hence, we are entitled to conclude
  that $\mathtt{Eff}_i(\mathtt{R}_0,\psi(\msg_0,\msg_1))$ is finite
  since $\mathtt{R}_0$ is so. It is now easy to see that
  $\mathsf{constrFP}(\{H_1(\knowzero_1)\},
  \mathtt{Eff}_i(\mathtt{R}_0,\psi(\msg_0,\msg_1))\cup
  \widehat{\mathit{Po}}(\know_1))$ terminates for reasons that are
  similar to those discussed for the computation of $\mathtt{R}_0$.
  The only difference is in the constraint Datalog rules derived from
  $\mathtt{Eff}(\mathtt{R}_0,\psi(\msg_0,\msg_1))$ for which it is not
  difficult to verify that the variables occurring in the body also
  occur in the head; thereby making it unnecessary to eliminate
  quantifiers as for $\mathtt{R}_0$. Thanks to basic properties of
  $\mathsf{constrFP}$ (see \cite{constraintdatalog}), we can derive that $\mathtt{R}_1$ is equal
  (modulo variable renaming) to $\mathcal{R}_1$.  By a straightforward
  induction, generalizing the previous observations on $\mathtt{R}_1$,
  it is possible to show that $\mathtt{R}_i$ is a finite set of
  constraint facts equal (modulo variable renaming) to
  $\mathcal{R}_i$, for $i\geq 2$.

  We are thus left with the problem of checking the
  $T_{\mathsf{S}}$-satisfiability of the (finite) set $\mathcal{R}_n$
  for constraint facts, for some $n\geq 0$.  This can be done as
  follows.  For each constraint fact $A \leftarrow \xi_i$ in
  $\mathcal{R}_n$, form the formula $\phi_A$ by taking the disjunction
  of all $\xi_i$'s for $i\geq 0$.  Then, take the disjunction of all
  the formulae $\phi_A$ built at the previous step and build the
  quantifier-free formula $\varphi$.  The satisfiability of the
  $\mathit{BSR}(T_{\mathsf{S}})$-formula $\varphi\wedge E$ is indeed decidable
  thanks to Theorem~\ref{th:dec-ext-BSR} in Section \ref{app:infprelim}. This concludes the proof.  
\end{proof}

\subsection{Main derivation of \CRO{} scenario}
\label{CR}

In this section, we illustrate the access control layer underlying (a
simplified version of) an e-Government application, first described
in~\cite{avantssar}. We already have a description of the \CRO{} case study in
the body of the paper (see Section~\ref{sec:runningexe}) and here we
show, step by step, the derivation process of its main access control query 
$\know(\CRep,\atoi(\Ed,\canstoredoc))$.

\begin{figure}[t] \center
 \subfigure[Informal view of state $s_3$]%
{\includegraphics[scale=.38]{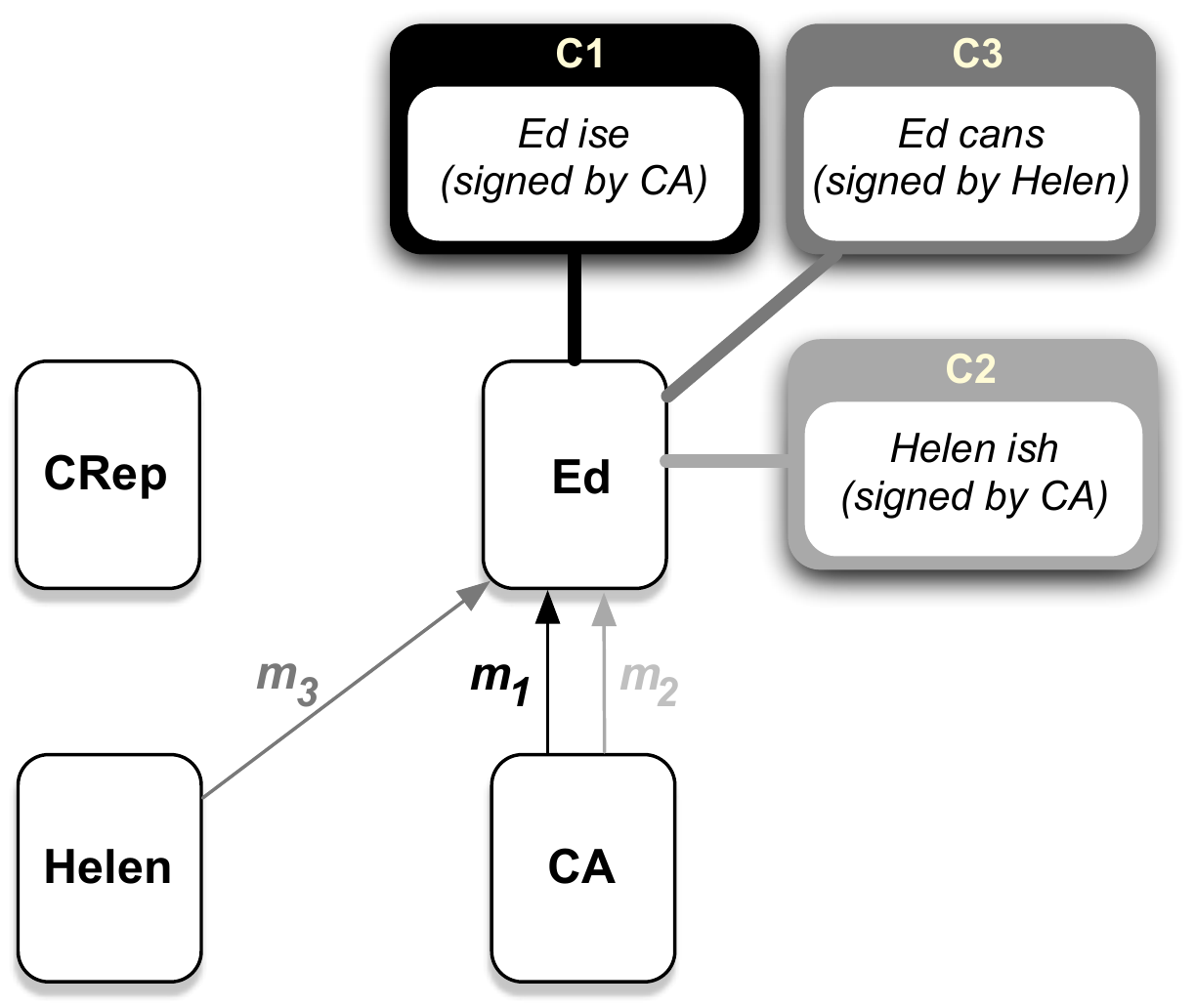}}\qquad\qquad\qquad
\subfigure[Certificate passing (first step) ]%
{\includegraphics[scale=.55]{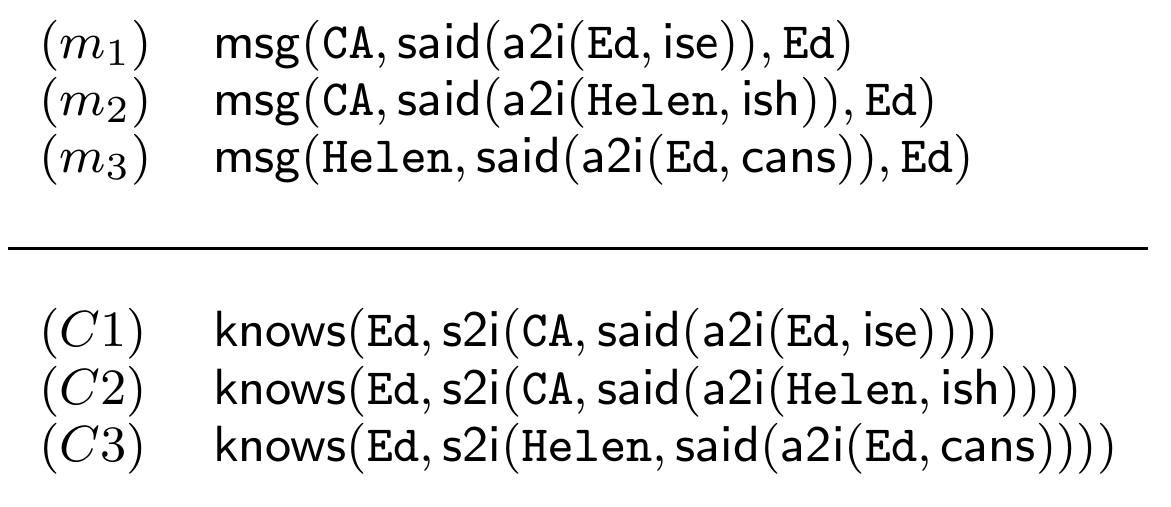}}\qquad\qquad 
\caption{\CRO{} certificate passing representation up to state $s_3$ \label{fig:CRO1}}
\end{figure}

A scenario of a possible run of the system is illustrated in
Fig.~\ref{fig:CRO1}(a) where, in the displayed state, the three
certificates ($C1$), ($C2$) and ($C3$) are in $\Ed$'s possession can be derived from the set of non-mechanizable facts as described in Example \ref{ex:certs} of Section \ref{subsec:exchangecerts}.\footnote{Some of the details of
the figure will be made clear below, e.g., the numbering of the states
and the fact that we only show the most interesting states of the
system.}
This is the result of the sending of three messages $m_1$, $m_2$, and $m_3$
(labeling the arrows in Fig.~\ref{fig:CRO1}(a)) as a consequence of the threefold application of the state-change rule \eqref{eq:send-action}.

After the successful processing of a car registration request, $\Ed$ is
willing to permanently store it in $\CRep$. In order to do this, he
should comply to the $\CRep$ policy that regulates access to its central
database. The decision of $\CRep$ to grant or deny to employees of a
\CRO{} the right to store a processed request in the database is based
on the rules ($P1$)--($P4$) described in Example \ref{ex:rules} of Section \ref{subsec:trustrelations} 
that we report again, informally, as follows: 
\begin{enumerate}
\item[($P1$)] an employee of a \CRO{} can store documents in the $\CRep$,
if the head of the \CRO{} permits it.
\end{enumerate}
Since the application of ($P1$) is based on valid certificates, further
policy regulations must specify the trust relationships that allow
$\CRep$ to validate the certificates in its possessions. Such rules are:
\begin{enumerate}
\item[($P2$)] certificates signed by $\theCA$ are trusted by anyone, 
\item[($P3$)] any certificate signed by $\theCA$ and countersigned by any
principal is trusted as one being signed by $\theCA$ itself, and 
\item[($P4$)] concerning the certificates about the permission of storing
documents, the head of a \CRO{} is trusted by anyone.
\end{enumerate}

In order to satisfy the access policies ($P1$)--($P4$) of $\CRep$, $\Ed$ is
supposed to forward to $\CRep$ the certificates 
in its possession, after signing each one of them. We want to remark that we preserve the countersign action
as described in the original scenario definition in \cite{avantssar}. As widely discussed in \cite{BRV-TR09},
the role of $\CRep$ is not purely passive but can potentially check the digital signatures of principals 
involved. Even if it is not relevant for our purposes in this paper, it is important to underline the capability of our formalization to model
more complex classes of scenarios.
After receiving $\Ed$'s certificates, $\CRep$ should be able to
grant him the right to store documents in its internal database using all the information
in its possession.



We want to define the \CRO{} scenario as an instance of the $\acs$ in Definition \ref{acss}.
So, we proceed defining each element of the transition system.

 Let $T_{\mathsf{S}}$ the effectively locally finite substrate theory underlying the \CRO{} 
as described in Section \ref{subsec:substh}, we take  $Const_{p}:=\{ \Ed, \Helen, \CRep,$ $\theCA\}$ and $Const_{a}:=\{ \isemployee, \ishead, \canstoredoc\}$ as the two (countably) sets of constants of sort $Principal$ and $Attribute$ to identify the four principals depicted in
Fig.~\ref{fig:CRO1}(a) and the attributes of being an employee, being an head and
having the right to store a document in the database, needed to built the initial certificates. 
The elements of $Const_{a}$ depend on the application we are considering and 
to characterize this set as particular primitive elements (not created by the ``function'' $\tdOn$) we have to add to the set $\mathit{In}$ the following axioms:
\begin{align*}
& \forall x. \mathsf{prim}(x) \rightarrow x=\isemployee \vee x=\ishead \vee x=\canstoredoc \quad \text{and}\\
& \isemployee \neq \ishead \wedge \isemployee \neq \canstoredoc \wedge \ishead \neq \canstoredoc
\end{align*}
as described in Section \ref{subsec:substh}.

We proceed with our formalization considering the set $K$ of $\{\knowzero\}$-atoms
and the set $M$ of $\{\msg\}$-atoms to represent the \textit{state} of the system according to Section \ref{subsec:setofstates}.
The initial situation is represented by the three non-mechanizable facts (as introduced in the Example \ref{ex:certsf}): 
\begin{align*}
(F1)\quad & \knowzero(\theCA, \atoi(\Ed,\isemployee)) \\
(F2)\quad & \knowzero(\theCA, \atoi(\Helen,\ishead))\\
(F3)\quad & \knowzero(\Helen, \atoi(\Ed,\canstoredoc)), 
\end{align*}
generated by the (arbitrary) human activities of $\theCA$ and $\Helen$, in order to put into the system the credentials needed to fulfill the established goal.
Considering the content of the network we define a state to be to be \textit{initial} if $M=\{\emptyset\}$.

Let $\mathit{Po}$ be the set of $\mathit{BSR}(T_S)$-formulae \eqref{eq:send-action}, \eqref{eq:internal-knowledge-is-knowledge}, \eqref{eq:receive-action}, together with axioms described according to Section \ref{subsec:substh} and application-dependent CLP rules ($P1$)-($P4$) of Example \ref{ex:rules},
%
where $p,q,$ and $r$ are variables of sort \mbox{Principal} and $x$ is a
variable of sort $\infon$. 
Clauses ($P2$), ($P3$) and ($P4$), stating the
trust relationship between the various principals, are required to
derive the hypotheses of ($P1$) in combination with the use of rule
\eqref{eq:trust-app}, as we will see below.

We have now all the information to define the \CRO{} case study as an instance of the $\acs$ according to Definition \ref{acss}. \\

  Let $\mathcal{CRO}=\langle \knowzero,\msg, I, \mathit{Po},
  \Psi\rangle$ be the \CRO{}-$\acs$ with substrate
  theory $T_{\mathsf{S}}$, where:
\begin{itemize}
\item $\know(\CRep,\atoi(\Ed, \canstoredoc))$ is the goal $G$ that must be satisfied, 
\item ($F1$), ($F2$), ($F3$) are the non-mechanizable facts declared above,
 \item $I(\msg)$ represents the first-order formula describing the initial state,
 \item $\mathit{Po}$ is the set of $\mathit{BSR}(T_S)$-formulae according to the definition above,
 \item $\Psi$ is a (finite) set of state-change rules of the form of (\ref{eq:change-rule-formula}).
\end{itemize}  
  
We can now easily describe the execution of the system representing and collecting the set $St$ of states,
by means of the two predicate symbols $\knowzero$ and $\msg$ which model the \textit{dynamic} part of the access control, unlike the \textit{static} one modeled by the first-order theory $T_{\mathsf{S}}$ previously defined. 

We will write $s:=K|M$ for a generic state, to represent the set $K \cup M$. 

\begin{figure}[t] 
  \centering
   \includegraphics[scale=.38]{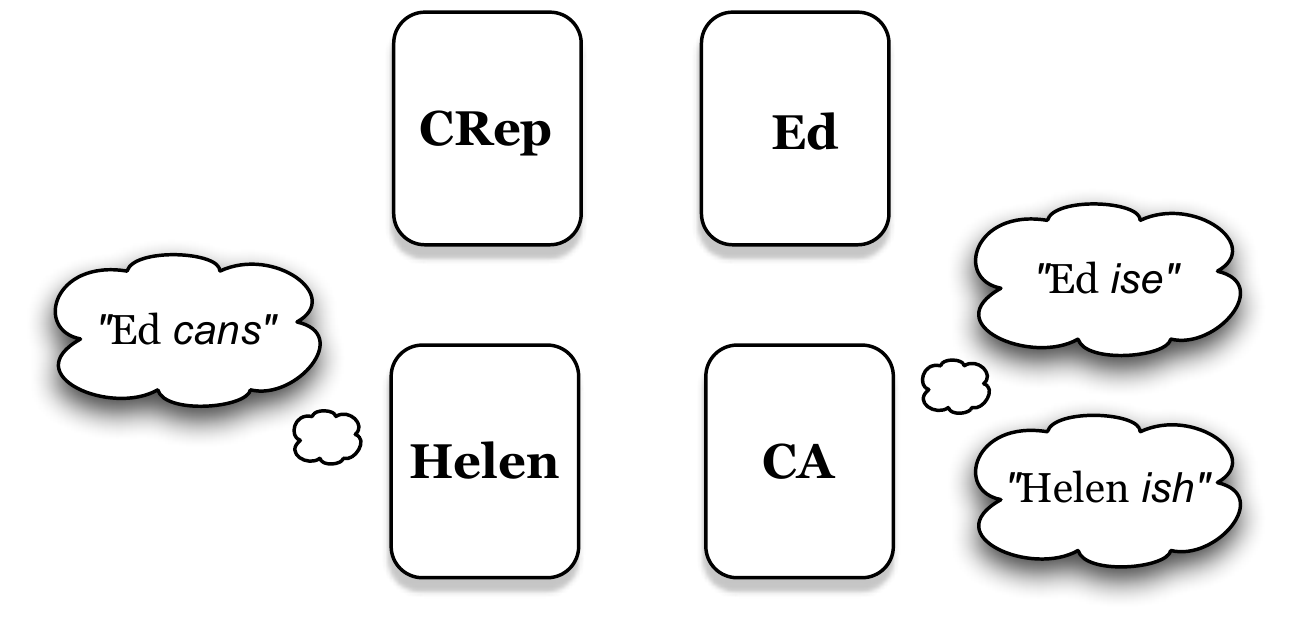}
  \caption{\CRO{} representation of the non-mechanizable facts  in the \textit{initial} state $s_0$ }
      \label{fig:CROinit}
\end{figure}

The initial situation in Fig.~\ref{fig:CROinit} (that for convenience we report here again) can be formalized by the following initial (symbolic) state:
  \begin{eqnarray*}
    s_0:= \left\{(F1), (F2), (F3)\right\} | 
    \emptyset.
  \end{eqnarray*}
 

Now, applying the rule (\ref{eq:internal-knowledge-is-knowledge}) in Section \ref{subsec:exchangecerts} and computing the fixed point by the $\mathsf{constrFP}$ procedure described in Fig. \ref{fig:fixpoint-comp}, 
we derive the following constraint facts: $\know(\theCA,
  \atoi(\Ed,\isemployee))$, $\know(\theCA, \atoi(\Helen,\ishead))$, and $\know(\Helen,
  \atoi(\Ed,\canstoredoc))$.

  
 At this point, we are ready to obtain the state $s_3$ depicted in Fig.~\ref{fig:CRO1}(a) by
repeatedly applying (three times) the state-change rule (\ref{eq:send-action}).
Considering the grounding substitution $\sigma_1:=\{ p\mapsto \theCA, q\mapsto \Ed,
  x\mapsto \atoi(\Ed,\isemployee) \}$, we have the
  following instance of (\ref{eq:send-action}):
  \begin{align*}
   \know(\theCA,\atoi(\Ed,\isemployee))~
     \Longrightarrow   \oplus \msg(\theCA, 
                        \said(\atoi(\Ed,\isemployee)),
                        \Ed) , 
  \end{align*}
  which is clearly enabled in $s_0$. The application's effect of $\psi\sigma_1$ on the constrained facts 
  just calculated in $s_0$ leads to
  \begin{eqnarray*}
    s_1:= K_0 | 
    \{ \msg(\theCA, \said(\atoi(\Ed,\isemployee)),
                        \Ed)  \},
  \end{eqnarray*}
  where $K_0=\{(F1), (F2), (F3)\}$.
  
It is not difficult to see
  that two further applications $\psi\sigma_2, \psi\sigma_3$ (where $\sigma_{1,2}$ are suitable ground substitutions) of
  (\ref{eq:send-action}) allow us to obtain state
  \begin{eqnarray*}
  &&  s_3:= K_0 | \left\{ 
    \begin{array}{l}
      \msg(\theCA, \said(\atoi(\Ed,\isemployee)), 
                        \Ed), \\
      \msg(\theCA, \said(\atoi(\Helen,\ishead)), 
                        \Ed), \\
      \msg(\Helen, \said(\atoi(\Ed,\canstoredoc)), 
                        \Ed), 
    \end{array}
    \right\},
  \end{eqnarray*}
  which is the formal counterpart of the configuration depicted in Fig.~\ref{fig:CRO1}.  
  
  It is also immediate to see that the repetitive application of the function $\mathsf{constrFP}$ in Fig.~\ref{fig:fixpoint-comp} to states $s_2$ and $s_3$ generates the following three
  facts (by repeatedly applying clause (\ref{eq:receive-action})):
 \begin{eqnarray*}
   && \know(\Ed, \stoi(\theCA, \said(\atoi(\Ed,\isemployee))))\,, \\ 
    &&\know(\Ed, \stoi(\theCA, \said(\atoi(\Helen,\ishead))))\,, \\ 
    && \know(\Ed, \stoi(\Helen, \said(\atoi(\Ed,\canstoredoc)))) 
  \end{eqnarray*}
 representing the formal counterpart of certificates ($C1$), ($C2$) and ($C3$) in Fig. \ref{fig:CRO1}(b).

\begin{figure}[t] \center
 \subfigure[Informal view of state $s_6$]%
{\includegraphics[scale=.38]{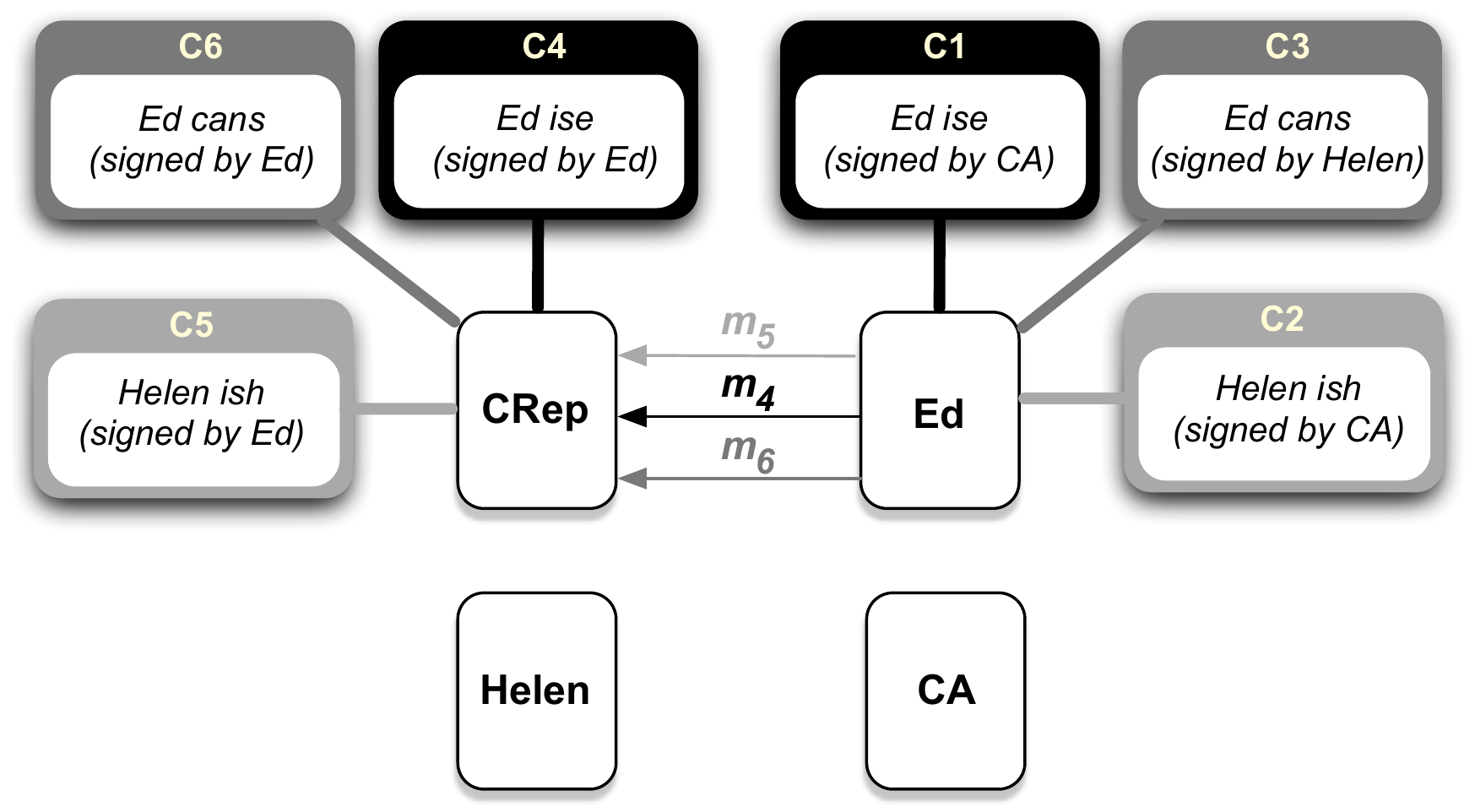}}\qquad\qquad\qquad
\subfigure[Certificate passing (second step)]%
{\includegraphics[scale=.55]{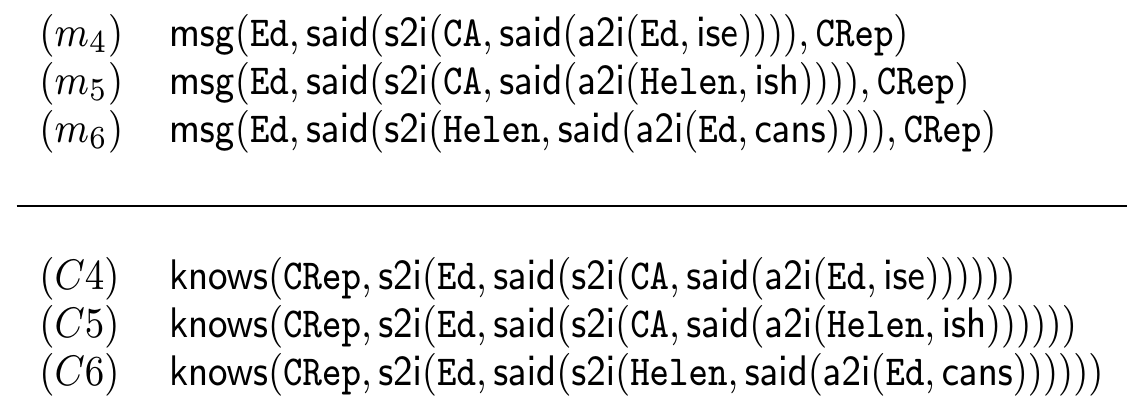}}\qquad\qquad 
\caption{\CRO{} certificate passing representation up to state $s_6$ \label{fig:CRO2}}
\end{figure}

Applying again the reasoning introduced up to now, 
these last three facts can be used by $\Ed$ to counter-sign the
  certificates and send them to $\CRep$ as depicted in
  Fig.~\ref{fig:CRO2}(a) (by appropriate instances $\psi\sigma_4, \psi\sigma_5, \psi\sigma_6$ of rule
  (\ref{eq:send-action}) with $\sigma_{4,5,6}$ suitable ground substitutions), thereby deriving the
  state
  \begin{align*}
  &  s_6:= K_0 ~|~ 
    M_3 ~\cup \\
   &   \left\{ 
    \begin{array}{l}
      \msg(\Ed,\said(\stoi(\theCA, \said(\atoi(\Ed,\isemployee)))),\CRep), \\
      \msg(\Ed,\said(\stoi(\theCA, \said(\atoi(\Helen, \ishead)))),\CRep), \\
      \msg(\Ed,\said(\stoi(\Helen, \said(\atoi(\Ed, \canstoredoc)))),\CRep) 
    \end{array}
    \right\},
  \end{align*} 
  which is the formalization of the configuration in Fig. \ref{fig:CRO2}, where $M_3$ abbreviates the second component of the state $s_3$ above. 
At the end of each single application of the state-change rule \eqref{eq:send-action}, the function $\mathsf{constrFP}$
returns (by using clause (\ref{eq:receive-action})) the set of certificates ($C4$),($C5$) and ($C6$) represented in Fig.\ref{fig:CRO2}(b). 

 Once the application has reached the state $s_6$, it is possible for $\CRep$ to
 take the decision to grant or deny to $\Ed$
the permission to store the processed request in the database.
To this end, we need to validate the certificates that are in possession of
$\CRep$ against the chain of trust relationships represented by the Horn
clauses ($P2$)--($P4$). More specifically, first, we consider the trust
relationships concerning the certificates about the roles of the
principals ($C1$), ($C2$) and then the certificate about the permission to store
documents in the database, ($C3$). Formally, this can be done by using the Horn
clause (\ref{eq:trust-app}) introduced in Section \ref{subsec:trustrelations}.  

In the
following, we describe which instances of (\ref{eq:trust-app}) need to
be considered and how their hypotheses are discharged.
In order to positively answer the query $G(\know(\CRep,
\atoi(\Ed, \canstoredoc)))$, we consider the following
instance of ($P1$): 
\begin{align*}
  (G) ~ \know(\CRep,\atoi(\Ed, \canstoredoc)) ~ &\leftarrow ~ \know(\CRep, \atoi(\Helen, \ishead)) \\
& \wedge ~ \know(\CRep, \atoi(\Ed, \isemployee)) \\
 &\wedge ~\know (\CRep, \stoi(\Helen,\said(\atoi(\Ed, \canstoredoc))) )))  ,
\end{align*}
let us call it ($G$) (to recall the goal of the reachability analysis problem of Section \ref{sec:DKAL-light}).  We have the problem of discharging the three hypotheses of ($G$), which can be grouped in two categories.  In fact,
the first two concern the roles of the principals (in particular, the
fact that $\Ed$ should be an employee and that $\Helen$ be the head of
the Car Registration Office), while the last is about the permission of
storing documents in the central repository.  We consider each
category in detail.  \\

\paragraph{Validation of certificates about the roles of principals}

Intuitively, we need to apply ($P3$) and ($P2$) so as to enable $\CRep$ to
derive the pieces of knowledge that $\Ed$ is an employee (fact ($H3$)
below) and that $\Helen$ is the head of the Car Registration Office
(fact ($H4$) below). Indeed, in the derivation, the certificates ($C4$) and
($C5$) will be used which, in turn, are obtained from ($C1$) and ($C2$) via
the applications of the state-change rule (\ref{eq:send-action})
described above. We begin by considering the following instances of
(\ref{eq:trust-app}):

\begin{small}
\begin{align*}
     (H1)~ & \know(\CRep,\stoi(\theCA, \said(\atoi(\Ed, \isemployee)))) ~ \leftarrow  \\
    & \know(\CRep,\stoi(\Ed,\said(\stoi(\theCA, \said(\atoi(\Ed, \isemployee)))))) ~ \wedge ~\\
    & \know(\CRep,\atoi(\Ed,\tdOn(\stoi(\theCA, \said(\atoi(\Ed, \isemployee))))))  \\
     & \\
    (H2)~ & \know(\CRep,\stoi(\theCA, \said(\atoi(\Helen, \ishead)))) ~ \leftarrow  \\
    & \know(\CRep,\stoi(\Ed,\said(\stoi(\theCA, \said(\atoi(\Helen, \ishead)))))) ~ \wedge \\
    & \know(\CRep,\atoi(\Ed,\tdOn(\stoi(\theCA, \said(\atoi(\Helen, \ishead))))))
\end{align*}
\end{small}

\noindent 
and notice that the first hypotheses are identical to ($C4$) and ($C5$)
respectively.  The second hypotheses of the two instances above are
identical to the following two instances of ($P3$) $\know(\CRep, \atoi(\Ed,\tdOn(\stoi(\theCA, \said(\atoi(\Ed, \isemployee))))))$ and 
$\know(\CRep, \atoi(\Ed,\tdOn(\stoi(\theCA, \said(\atoi(\Helen, \ishead))))))$.  
So, we are entitled to consider the heads of the two instances of
(\ref{eq:trust-app}) above as derived ground facts; let us call them
($H1$) and ($H2$), respectively.  Then, consider two more instances of
(\ref{eq:trust-app}):
\begin{align*}
   (H3) ~& \know(\CRep,\atoi(\Ed, \isemployee)) ~ \leftarrow \\
   &  \know(\CRep,\stoi(\theCA,\said(\atoi(\Ed, \isemployee)))) ~ \wedge ~ \\
   & \know(\CRep,\atoi(\theCA,\tdOn(\atoi(\Ed, \isemployee))))  \\
   & \\
    (H4) ~ &\know(\CRep,\atoi(\Helen, \ishead)) ~ \leftarrow \\
   &  \know(\CRep,\stoi(\theCA,\said(\atoi(\Helen, \ishead)))) ~ \wedge ~ \\
   &  \know(\CRep,\atoi(\theCA,\tdOn(\atoi(\Helen, \ishead)))) 
\end{align*}

The first hypotheses of these two ground Horn clauses are identical to
($H1$) and ($H2$) respectively. 

Their second hypotheses are
identical to the following two instances of ($P2$) $\know(\CRep, \atoi(\theCA, \tdOn(\atoi(\Ed, \isemployee))))$ and
$\know(\CRep, \atoi(\theCA, \tdOn(\atoi(\Helen, \ishead))))$.
As a consequence, we can consider the heads of the last two instances
of (\ref{eq:trust-app}) as derived ground facts; let us call them 
($H3$) and ($H4$), respectively. 

\paragraph{Validation of certificates about the permission of storing
  documents}  Intuitively, we need to apply ($P4$) so as to make
immediately available to $\CRep$ the knowledge about the authorization
state concerning the fact that $\Ed$ is permitted to store documents
in the central repository by $\Helen$ (fact ($H6$) below).
We begin by considering the following instance of ($P4$):

\begin{small}
\begin{align*}
    (H5) ~ & \know(\CRep, \atoi(\Ed, \tdOn(\stoi(\Helen, \said(\atoi(\Ed, \canstoredoc)))))) \\
    & \leftarrow \know(\CRep, \atoi(\Helen, \ishead)).
\end{align*}
\end{small}

\noindent
Since the hypothesis of this Horn clause is identical to ($H4$), we are
entitled to consider the head of this clause as a derived fact; let us
call it ($H5$).  Then, consider the following instance of
\eqref{eq:trust-app}:

\begin{small}
\begin{align*}
   (H6)~ & \know(\CRep,\stoi(\Helen, \said(\atoi(\Ed, \canstoredoc)))) ~ \leftarrow  \\
   & \know(\CRep,\stoi(\Ed,\said(\stoi(\Helen, \said(\atoi(\Ed, \canstoredoc)))))) ~ \wedge ~\\
    & \know(\CRep,\atoi(\Ed,\tdOn(\stoi	(\Helen, \said(\atoi(\Ed, \canstoredoc)))))) .
\end{align*}
\end{small}

\noindent
The first hypothesis of the instance is identical to ($C6$) and the
second hypothesis is equal to ($H5$); thus, we can consider its head as
a derived ground fact, let us call it ($H6$).

\paragraph{Putting things together}  At this point, it is sufficient to
observe that the first hypothesis of ($G$) is ($H4$), the second is ($H3$),
and the last is ($H6$) so that we can consider the head of the rule as a
derived ground fact which is precisely the query that we were
interested to answer, $\know(\CRep,\atoi(\Ed, \canstoredoc))$.\\

As a final remark, we observe that it is possible to model several
scenarios by considering alternative ways of distributing the
certificates among the principals.  For example, initially, only the
certificate about his role can be sent to $\Ed$, that concerning the
role of $\Helen$ is sent to her, and that about the permission for
$\Ed$ to store documents in the central repository can be sent to
$\CRep$ directly from $\Helen$.  Indeed, this changes the way in which
we can derive the query of interest as certificates are counter-signed
by different principals so that trust relationships must be chained
differently.

\subsection{Dkal implementation for \CRO{} scenario}
In this section, we give a concrete implementation of the proposed CRO scenario in Section \ref{sec:runningexe},
by using the DKAL distributed authorization policy language provided by Microsoft Research \cite{msr}. 
The DKAL project page \cite{dkalimpl} contains a downloadable engine (implemented in F$\#$) for running and checking DKAL policies. 
We implemented and successfully tested the scenario proposed in this paper, and we give the corresponding code in the following:\\

\hrule
\vspace*{.2cm}
\textbf{Input specification code}
\vspace*{.2cm}
\hrule
\small
\begin{verbatim}
type Principal = Dkal.Principal
type Infon = Dkal.Infon
type Attribute = System.String
type Evidence = Dkal.Evidence

relation hasrole(P: Principal, A: Attribute)
relation haspermission(P: Principal, A: Attribute)

---crep------------
//crep's policy
//rule to learn every justified infon comes from the communication
with X: Infon, E: Evidence
upon X [E]
do learn X [E]

//(P1) rule
with P: Principal, Q: Principal, R: Principal, 
E: Evidence, E1: Evidence, E2: Evidence
if hasrole(P, ish) [E]
  if hasrole(Q, ise) [E1]
    if R said haspermission(Q, cans) [E2]
do once send to ed: haspermission(P, "ok ed, u can write!")

//(P2) rule	
with P: Principal, X: Infon, E: Evidence
upon theca said X [E]
do learn  X

//(P3) rule
with P: Principal, X: Infon, E: Evidence
upon P said theca said X [E]
do learn theca said X

//(P4) rule
with P: Principal, Q: Principal, R: Principal, E: Evidence, E1: Evidence 
if hasrole(P, ish) [E]
learn hasrole(P, ish) -> Q said P said hasrole(R, ise) [E1] 

---ed----------------
//ed's policy
with P: Principal, Q: Principal, R: Principal,  A: Attribute, E: Evidence, X: Infon
upon X -> R said haspermission(P, A) [E]
do once send  to crep: Me said haspermission(Me, A) [E]

with Q: Principal, R: Principal, A: Attribute, E: Evidence, X: Infon
upon Q said hasrole(R, A) [E]
do once say with justification to crep: Q said hasrole(R, A) [E]

---theca-------------
//theca's policy
// internal knowledge about roles of principals

substrate xml("<roleAssignments>
	<roleAssignment id='ed' role='ise' />
	<roleAssignment id='helen' role='ish' />
	</roleAssignments>")
namespaces "roleAssignments"
		
with P: Principal, P1: Principal, A: Attribute, A1: Attribute
if  asInfon({| "roleAssignments" | "//roleAssignment[@role='ise']/@id" | P |}) &&
asInfon({| "roleAssignments" | "//roleAssignment[@id='ed']/@role" | A |}) &&
asInfon({| "roleAssignments" | "//roleAssignment[@role='ish']/@id" | P1 |}) &&
asInfon({| "roleAssignments" | "//roleAssignment[@id='helen']/@role" | A1 |}) 
do say with justification to P: hasrole(P, A)
	   say with justification to P: hasrole(P1, A1)
		
---helen----------------
//helen's policy
// internal knowledge about permissions of principals

substrate xml("<permissionAssignments>
	<permissionAssignment id='ed' perm='cans' />
	</permissionAssignments>")
namespaces "permissionAssignments"

with P: Principal, A: Attribute
if asInfon({| "permissionAssignments" | "//permissionAssignment[@perm='cans']/@id" | P |}) &&
asInfon({| "permissionAssignments" | "//permissionAssignment[@id='ed']/@perm" | A |})
do send with justification to P:
	// delegation to Ed to say haspermission condition must be entailed by Ed's 
	// infostrate
	
	with Q: Principal, A1: Attribute
	Q said haspermission(Q, A1) -> Me said haspermission(P, A)	
	\end{verbatim}
	\hrule
\vspace*{.2cm}
\textbf{Output specification code}
\vspace*{.2cm}
\hrule
\small
\begin{verbatim}
>> From theca to ed:
theca said hasrole(ed, "ise") [ signed by theca 1693769001 ] &&
theca said hasrole(helen, "ish") [ signed by theca 1939556616 ]
 
>> From helen to ed:
with Q: Dkal.Principal, A1: System.String
    Q said haspermission(Q, A1) ->
    helen said haspermission(ed, "cans")
 [ signed by helen -973755704 ]
 
>> From ed to crep:
ed said haspermission(ed, "cans") [ signed by helen -973755704 ] &&
ed said theca said hasrole(ed, "ise") [ signed by theca 1693769001 ] 
                                      [ signed by ed -855343925 ] &&
ed said theca said hasrole(helen, "ish") [ signed by theca 1939556616 ] 
                                         [ signed by ed -14984535 ]
 
>> From crep to ed
ok ed, u can write! 
 
Fixed-point reached
\end{verbatim}

\end{document}